\documentclass[aps,amsmath,amssymb,10pt,floatfix,twocolumn,showpacs,amfonts,floatfix,superscriptaddress,longbibliography,reprint]{revtex4-2}

\pdfoutput=1
\usepackage{amsthm,amsmath,amssymb,amsfonts,bbm} 
\usepackage{mathrsfs} 
\usepackage{lipsum}

\usepackage{graphicx} 
\usepackage{subfigure}
\usepackage{ragged2e}
\usepackage{xcolor}
\usepackage[colorlinks=true,citecolor=blue,urlcolor=blue,linkcolor=red,hyperindex]{hyperref}

\usepackage{tikz,tikz-3dplot,tikz-dependency}
\usetikzlibrary{quotes,angles,arrows,arrows.meta,positioning,calc,snakes,fadings,shadings,intersections,decorations.text}
\usepackage{overpic}

\usepackage{amsmath,bm}
\usepackage{natbib}
\usepackage{siunitx}
\usepackage{witharrows}
\usepackage[title]{appendix}

\begin{document}

\title{A spin-rotation mechanism of Einstein–de Haas effect based on a ferromagnetic disk}

\author{Xin Nie}
\author{Jun Li}
\affiliation{Guangdong Provincial Key Laboratory of Magnetoelectric Physics and Devices, State Key Laboratory of Optoelectronic Materials and Technologies, Center for Neutron Science and Technology, School of Physics, Sun Yat-Sen University, Guangzhou, 510275, China}

\author{Trinanjan Datta}
\email{tdatta@augusta.edu}
\affiliation{Department of Physics and Biophysics, Augusta University, 1120 15th Street, Augusta, Georgia 30912, USA}
\affiliation{Kavli Institute for Theoretical Physics, University of California, Santa Barbara, California 93106, USA}

\author{Dao-Xin Yao}
\email{yaodaox@mail.sysu.edu.cn}
\affiliation{Guangdong Provincial Key Laboratory of Magnetoelectric Physics and Devices, State Key Laboratory of Optoelectronic Materials and Technologies, Center for Neutron Science and Technology, School of Physics, Sun Yat-Sen University, Guangzhou, 510275, China}
\affiliation{International Quantum Academy, Shenzhen 518048, China}

\begin{abstract}
Spin-rotation coupling (SRC) is a fundamental interaction that connects electronic spins with the rotational motion of a medium. We elucidate the Einstein-De Haas (EdH) effect and its inverse with SRC as the microscopic mechanism using the dynamic spin-lattice equations derived by elasticity theory and Lagrangian formalism. By applying the coupling equations to an iron disk in a magnetic field, we exhibit the transfer of angular momentum and energy between spins and lattice, with or without damping. The timescale of the angular momentum transfer from spins to the entire lattice is estimated by our theory to be on the order of 0.01 ns, for the disk with a radius of 100 nm. Moreover, we discover a linear relationship between the magnetic field strength and the rotation frequency, which is also enhanced by a higher ratio of Young's modulus to Poisson's coefficient. In the presence of damping, we notice that the spin-lattice relaxation time is nearly inversely proportional to the magnetic field. Our explorations will contribute to a better understanding of the EdH effect and provide valuable insights for magneto-mechanical manufacturing.
\end{abstract}

\date{\today}

\maketitle
\section{Introduction}
\begin{figure*}[t!]
	\centering
	\includegraphics[width = 1.7 \columnwidth]{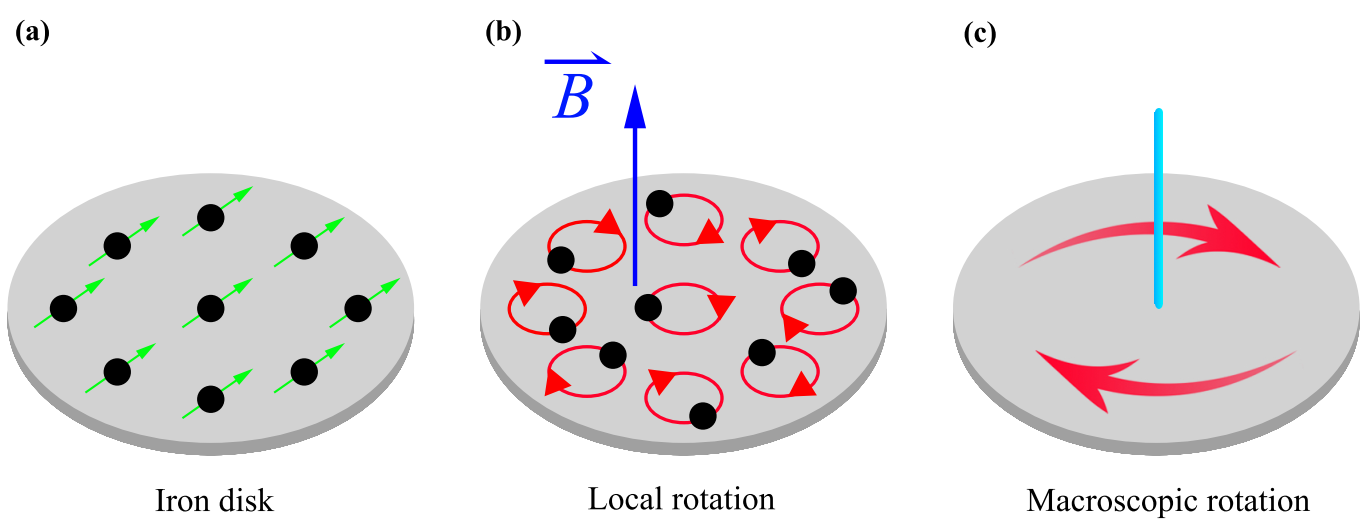}
	\caption{Schematic diagram of SRC mechanism in an iron disk. (a) Initial ferromagnetic configuration. (b) After the application of an external magnetic field $\bm{B}$ in the $z$-direction, the change in spin angular momentum causes atomic motion, resulting in local rotation of the disk. (c) The microscopic rotational "message" is transmitted throughout the disk at the speed of sound, leading to the macroscopic EdH rotation, as discussed in Refs. \cite{Losby_2018, tauchert2022polarized}.}\label{fig:mechanism}
\end{figure*}
Magnetization-rotation coupling is an intriguing and long-lasting topic. Richardson \cite{PhysRevSeriesI.26.248} first proposed that the moment of momentum is proportional to the magnetization using Ampère's molecular currents. Soon after, the gyromagnetic effect in macroscopic bodies was observed in the EdH experiment \cite{einstein1915experimenteller} (change of magnetization induces mechanical rotation) and the inverse Barnett experiment \cite{barnett1915magnetization} (mechanical rotation triggers magnetization). In that period, molecular orbital theory was used to explain the EdH effect---the external magnetic field affected the angular momentum generated by electron orbital motion, leading the iron cylinder in the experiment to produce a mechanical angular momentum \cite{einstein1915experimental}. However, the discovery of electron spin reveals that the gyromagnetic ratio (ratio of magnetic moment to angular momentum) measured experimentally is close to the gyromagnetic ratio of the spin, $e/m$, which is twice the value predicted by considering only electron orbital motion \cite{RevModPhys.34.102}. It is now widely recognized that spin is an essential origin of magnetism and that the spin-lattice coupling is indispensable for the gyromagnetic effect. For transition metals like iron, the electronic orbital angular momentum can be readily frozen by the surrounding crystal field, resulting in a dominant contribution of spin to the magnetic moment. In a contemporary interpretation of the EdH effect, the conservation of total angular momentum is a basic principle, i.e., any change in the spin angular momentum requires a corresponding compensation of the mechanical angular momentum.

At the microscopic level, the mechanism via which angular momentum is transferred from electrons to the entire body remains elusive, as it is insufficient to explain the EdH effect by the conservation of angular momentum alone. Researchers firstly pay attention to those EdH effects occurring at the molecular scale due to finite degrees of freedom. For instance, the EdH effect is studied for a system of two dysprosium atoms trapped in a spherically symmetric harmonic potential \cite{Gorecki_2016}. A noncollinear tight-binding model capable of simulating the EdH effect of an $\mathrm{O_2}$ dimer has been proposed \cite{10.1063/1.5092223}. With the development of molecular-spintronics \cite{doi:10.1126/science.1065389, xiong2004giant, sanvito2006molecular, bogani2008molecular}, the problem of single spin embedded within a macroscopic object has gained attention \cite{chudnovsky2005universal, PhysRevB.81.214423, PhysRevLett.72.3433}. The concept of phonon spin is used to explicate the spin-phonon coupling involved \cite{PhysRevLett.112.085503}. The orbital part and the spin part of phonon angular momentum (now considered pseudo-angular momentum \cite{PhysRevB.103.L100409}) are clearly divided in Refs. \cite{PhysRevB.97.174403} and \cite{PhysRevB.92.024421}, where their exchange with (real) spin angular momentum are also discussed. Experiments have progressed to realize the coupling of single-molecule magnets with nanomechanical resonators \cite{ganzhorn2016quantum}. Nevertheless, the multiatomic EdH effect, which involves numerous degrees of freedom, is inadequately comprehended despite the use of molecular dynamics and micromagnetic simulations \cite{AMANN2019217, PhysRevB.93.060402, PhysRevB.78.024434, DEDNAM2022111359}. The mechanical dynamics of a magnetic cantilever resulting from changes in magnetization has also been described \cite{10.1063/1.2355445, PhysRevB.79.104410, PhysRevB.89.174420}, but only one-sidedly considering the effect of spin evolution on lattice mechanics. Therefore, it is desired to develop a method that treats spin and lattice as equally important and converts the microscopic mechanism of the EdH effect into a macroscopic gyroscopic effect. 

In this paper, we utilize Lagrangian formalism in classical field theory to derive spin-lattice dynamical equations in an elastic ferromagnet, where spins are coupled by exchange interactions. We assume that the magnetism of iron atom originates from the electronic spin, $\bm{M}_{i}=\gamma\hbar\bm{S}_{i}$, where $\bm{M}_{i}$ is magnetic moment of atom $i$, $\hbar$ is the Planck's constant, $\gamma=ge/(2m_{e})$ is the gyromagnetic ratio, and $g$ factor is 2 for spin $\bm{S}_{i}$. By means of the spin-rotation Hamiltonian arising from the conservation of spin angular momentum and mechanical angular momentum, a theoretical framework for the spin-lattice coupling is constructed---the change of magnetization results in a force on the atom, whereas the evolution of atomic displacement produces a torque acting on the spin. Then, we numerically solve the spin-rotation dynamical equations in an iron disk to reveal the transfers of angular momentum and energy, thereby validating the EdH effect. The characteristics exhibited during the coupling process, such as the transfer timescale and the dependence of the rotation frequency of the system on magnetic field and material parameters, can enhance the comprehension for laser-induced ultrafast demagnetization in solids \cite{PhysRevLett.76.4250, PhysRevB.78.174422, fahnle2018ultrafast, dornes2019ultrafast}. Exploring the mutual transfer of mechanical angular momentum and spin angular momentum can aid in creating new mechanical techniques for manipulating electron spins, or alternatively, using spins to manipulate mechanical motion. 

This work is organized as follows. The spin-rotation Hamiltonian and dynamics equations merging spin and lattice are shown in Section \ref{Section:spin-rotation dynamics}. The numerical method of applying the coupling equations to a disk model is introduced in Section \ref{Section:numerical solution}. And in Section \ref{Section:image results}, the pictures of angular momentum and energy transfers with and without damping are presented. The effects of the magnetic field on the rotation frequency and spin-relaxation time of the system are also investigated. And the Barnett effect is briefly discussed through the SRC mechanism. In Section \ref{Section:conclusions}, we draw conclusions and propose some further promising research ideas on spin-lattice coupling.

\section{spin-rotation Hamiltonian and coupling dynamical equations }\label{Section:spin-rotation dynamics}
Recently, a study using ultrafast electron diffraction to probe lattice dynamics after femtosecond laser excitation has captured the transfer of angular momentum from spins to the atoms, which induce circularly polarized phonons as the atoms rotate around their equilibrium positions \cite{tauchert2022polarized}. These findings may shed light on the microscopic mechanism underlying the EdH effect. For a stationary elastic magnet, if it is stimulated by an external field like a magnetic field or heat bath, the spins within the magnet will change their orientation. It is assumed here that the spins are formed by strong exchange interactions that only alter their direction but not the absolute value. As the system relaxes back to equilibrium, the magnet generates mechanical rotation locally with atoms rotating around their equilibrium positions, to preserve the total angular momentum of the system. Subsequently, this local rotation transitions into an overall rotation (i.e. the EdH macroscopic rotation), where all the atoms synchronously rotate around the center of the lattice \cite{Losby_2018}. This interaction between spins and lattice is called SRC, and a visual representation of it is depicted in Fig. \ref{fig:mechanism}.

Consider a magnetic system characterized by the magnetic moment $\bm{{M}}$ and the mechanical angular momentum $\bm{{L}}$. Evidently, the magnetic moment $\bm{{M}}$ corresponds to an internal angular momentum, $\gamma^{-1}\bm{{M}}$. In the absence of external moments, $\bm{{L}}$ and $\gamma^{-1}\bm{{M}}$ should satisfy the conservation of total angular momentum,
\begin{equation}\label{angular_const}
\bm{{L}}+\gamma^{-1}\bm{{M}}=\mathrm{{constant}}.
\end{equation}
Taking the time derivative on both sides of Eq. (\ref{angular_const}) yields $\dot{\bm{L}}+\gamma^{-1}\dot{\bm{{M}}}=0$, where $\dot{\bm{L}}$ represents the rate of change of mechanical angular momentum. The change in mechanical angular momentum indicates that a mechanical torque must have been generated,
\begin{equation}\label{torque}
\bm{{T}}=-\gamma^{-1}\dot{\bm{{M}}}.
\end{equation}
Eq. (\ref{torque}) informs us that a change in magnetization produces a mechanical torque that rotates magnetic atoms. By denoting $\bm{\phi}_{j}$ as the rotation angle at site $j$ and
considering the relationship $\bm{M}_{i}=\gamma\hbar\bm{S}_{i}$, the spin-rotation Hamiltonian in the lattice can be generally expressed as
\begin{equation}\label{eq:coupling_Hamiltonian2}
H_\mathrm{s-r}=-\sum_{ij}A_{ij}^{ab}\hbar\dot{S}_{i}^{a}\phi_{j}^{b}.
\end{equation}
Here, $A_{ij}^{ab}$ is the coupling coefficient with $a,b\in{x,y,z}$. 

Next, we introduce other interactions in the system. The spin $\boldsymbol{S}_{i}$ at $i$th site can be decomposed into two mutually perpendicular components,
\begin{equation}
\boldsymbol{S}_{i}=\boldsymbol{\mu}_{i}+\boldsymbol{n}_{i}\sqrt{S^2-\mu_{i}^2},
\end{equation}
where $\boldsymbol{\mu}_{i}\cdot\boldsymbol{n}_{i}=0$, but the orientations of themselves are arbitrary. $S$ is the size of the spin. In terms of $\boldsymbol{\mu}_{i}$, $\boldsymbol{n}_{i}$ and  $\boldsymbol{u}_{i}$ (the displacement of the $i$th atom), the kinetic energy of the system can be written as \cite{landau2013classical, PhysRevLett.129.167202} \begin{equation}\label{eq:kinetic_energy}
T=\frac{\hbar}{2S}\sum_{i=1}(\dot{\boldsymbol{\mu}}_{i}\times\boldsymbol{n}_{i})\cdot\boldsymbol{\mu}_{i}+\sum_{i=1}\frac{1}{2}m_{i}\dot{\boldsymbol{u}}^{2}_{i},
\end{equation}
with $m_{i}$ the mass of the $i$th atom. The first term represents the spin kinetic energy and the second term represents the lattice kinetic energy.

The total potential energy contains the Heisenberg exchange interaction $H_\mathrm{ex}$, the external Zeeman energy $H_{Z}$ and the crystal elastic energy $H_{\mathrm{e}}$ \cite{reddy2006theory},
 \begin{align}
H_\mathrm{ex}&=-\frac{1}{2}\sum_{ij}I_{ij}\boldsymbol{S}_{i}\cdot\boldsymbol{S}_{j}, \label{eq:Heisenberg_exchange}\\
H_{Z}&=\sum_{i}g\mu_{B}\boldsymbol{B}\cdot\boldsymbol{S}_{i},\label{eq:Zeeman}\\
H_{\mathrm{e}}&=\frac{1}{2}\int d^3 r~ \sigma_{\alpha\beta}u_{\alpha\beta}.\label{eq:Elastic}
\end{align}
Here $I_{ij}$ is the ferromagnetic exchange coupling between $i$th and $j$th sites, $\mu_{B}$ is the Bohr magneton, $\bm{B}$ is the external magnetic
field, and $\sigma_{\alpha\beta}$, $u_{\alpha\beta}$ are respectively the $(\alpha,\beta)$ components of the stress tensor and strain tensor with directional indexes $\alpha$, $\beta$. 

Considering the large difference between the system size and interatomic distance, as well as the relatively low-energy associated with the transfer in comparison to the overall energy of the system, we treat the medium as continuous and employ the displacement field $\bm{u}(\bm{r},t)$ and the spin field $\bm{S}(\bm{r},t)$ at position $\bm{r}$ and time $t$ to describe the degrees of freedom of lattice and spin, respectively. For linearly elastic materials, the relationship between $\sigma_{\alpha\beta}$ and $u_{\alpha\beta}$ conforms to Hooke's law, where $u_{\alpha\beta}$ is given by $u_{\alpha\beta}=(\nabla_{\alpha}u_{\beta}+\nabla_{\beta}u_{\alpha})/2$, and $\sigma_{\alpha\beta}$ can be expressed as $\sigma_{\alpha\beta}=C_{\alpha\beta\gamma\delta}u_{\gamma\delta}$. Here, $C_{\alpha\beta\gamma\delta}
=\mathcal{R}_{~\alpha}^{\mu}\mathcal{R}_{~\beta}^{\nu}\mathcal{R}_{~\gamma}^{\sigma}\mathcal{R}_{~\delta}^{\rho}C_{\mu\nu\sigma\rho}$, with the coefficient tensor $C$, rotation transformation matrix $\mathcal{R}$, and directional indexes $\gamma$, $\delta$. 

We believe that local coupling accounts for the main contribution \cite{PhysRevB.79.104410}, so this study focuses on the local coupling. According to the elastic theory, $\bm{\phi}$ can be written as $\bm{\phi}(\boldsymbol{r},t)=\nabla\times\boldsymbol{u}(\boldsymbol{r},t)/2$ \cite{landau1986theory}. We introduce the spin density $\boldsymbol{S}^{\prime}$, defined as $\boldsymbol{S}^{\prime}=\boldsymbol{S}/V$, where $V$ is the volume of the primitive cell and $\bm{S}$ is the total spin within $V$. For simplicity, we assume that the SRC in the plane is isotropic. Consequently, the continuous form of $H_\mathrm{s-r}$ can be given as
\begin{align}\label{eq:Coupling}
H_\mathrm{s-r}=\frac{1}{2}\int d^3 r~\hbar\dot{\boldsymbol{S}}^{\prime}\cdot(\nabla\times\boldsymbol{u}).
\end{align}
Eq. (\ref{eq:Coupling}) incorporates the spins and the rotational motion of the lattice, enabling the exchanges of angular momentum and energy between them.

For the nearest neighbor Heisenberg term, we approximate it by the Taylor expansion, whereby $\boldsymbol{S}_{j}=\boldsymbol{S}_{i+\delta_{j}}=\boldsymbol{S}_{i}+(\boldsymbol{\delta}_{j}\cdot\nabla)\boldsymbol{S}_{i}+\frac{1}{2}(\boldsymbol{\delta}_{j}\cdot\nabla)^2\boldsymbol{S}_{i}$ and $\boldsymbol{\delta}_{j}=\boldsymbol{S}_{j}-\boldsymbol{S}_{i}$. Since the Taylor expansion form is restricted by the high rotational symmetry, other lattice choices introduce nothing but the coefficient difference, which can be absorbed further into parameter selection, thus we take the square lattice as an example. In order to correctly obtain the spin dynamics equation, technically, we fix the direction of $\bm{n}$, $\partial \bm{n}/{\partial t}= 0$. The disk material is considered isotropic with $I_{ij}=I$ for simplicity. We replace $\boldsymbol{S}_{i}$ and $\bm{u}_{i}$ with $\boldsymbol{S}(\bm{r},t)$ and $\bm{u}(\bm{r},t)$, respectively, and introduce the definitions $\boldsymbol{\mu}^{\prime}=\boldsymbol{\mu}/V$, $S^{\prime}=S/V$ and $I^{\prime}=I\times V$. Under Eqs. (\ref{eq:kinetic_energy})-(\ref{eq:Elastic}), (\ref{eq:Coupling}), the Lagrangian density of the system reads,
\begin{align}\label{eq:Lagrangian density}
\mathcal{L}=&\frac{\hbar}{2S^{\prime}}(\dot{\boldsymbol{S}^{\prime}}\times\boldsymbol{n})\cdot\boldsymbol{S}^{\prime}+\frac{1}{2}\rho\dot{\boldsymbol{u}}^{2}+\frac{I^{\prime}}{2}\boldsymbol{S}^{\prime}\cdot[\boldsymbol{S}^{\prime}+a^2\nabla^2\boldsymbol{S}^{\prime}]\nonumber \\&-g\mu_{B}\boldsymbol{B}\cdot\boldsymbol{S}^{\prime}-\frac{1}{2}C_{\alpha\beta\gamma\rho}u_{\gamma\rho}u_{\alpha\beta}-\frac{1}{2}\hbar\dot{\boldsymbol{S}}^{\prime}\cdot(\nabla\times\boldsymbol{u}),
\end{align}
where $\rho$ is the mass density of the material and $a$ is the lattice constant. Based on the Euler-Lagrange equations in classical field theory,
\begin{subequations}
	\begin{gather}
	\frac{d}{dt}\frac{\partial \mathcal{L}}{\partial \dot{u}_{\alpha}}+\nabla\cdot\frac{\partial \mathcal{L}}{\partial (\nabla u_{\alpha})}=\frac{\partial \mathcal{L}}{\partial u_{\alpha}}, \\
	\frac{d}{dt}\frac{\partial \mathcal{L}}{\partial \dot{S}_{\alpha}^{\prime}}+\nabla\cdot\frac{\partial \mathcal{L}}{\partial (\nabla S_{\alpha}^{\prime})}=\frac{\partial \mathcal{L}}{\partial S_{\alpha}^{\prime}},
	\end{gather}
\end{subequations}
with $\alpha =x$, $y$, and $z$, the spin-lattice dynamic equations can be derived,
\begin{gather} 
\rho\frac{\partial^2 \bm{u}}{\partial t^2}-\nabla\cdot\bm{\sigma}+\frac{1}{2}\nabla\times\hbar\dot{\bm{S}^{\prime}}=0, \label{eq:displacement}\\
\hbar\dot{\bm{S}}=\bm{S}\times \left[Ia^2\nabla^2 \bm{S}-g\mu_{B}\bm{B}+\frac{\hbar}{2}(\nabla\times\dot{\bm{u}})\right]. \label{eq:spin vector dynamics}
\end{gather}
Eqs. (\ref{eq:displacement}) and (\ref{eq:spin vector dynamics}) tell us that the time evolution of the spin field acts as a driving force $\boldsymbol{f}^{(R)}=-(\nabla\times\hbar\dot{\boldsymbol{S}}^{\prime})/2$ on the displacement field, and in turn, the evolution of the displacement field produces a torque $\hbar(\nabla\times\dot{\bm{u}})/2$ acting on the spin. Note that the entire rotation contributes to $\nabla\times\dot{\bm{u}}$.

Let us now turn to angular momentum. By previous discussion [see Eq. (\ref{angular_const})], the total angular momentum (spin and mechanical) can be represented as
\begin{equation}\label{eq:angular momentum}
\boldsymbol{J}=\int d^3r ~ [\hbar\boldsymbol{S}^{\prime}+\rho(\boldsymbol{r}\times\dot{\boldsymbol{u}})].
\end{equation}
For the time derivative of the $\alpha$-component of $\boldsymbol{J}$,
\begin{equation}\label{eq:angular momentum derivative}
\dot{J}_{\alpha}=\int d^3r ~  \hbar\dot{S}_{\alpha}^{\prime}+\int d^3 r ~ \epsilon_{\alpha\beta\gamma}r_{\beta}\rho\ddot{u}_{\gamma},
\end{equation}
the following expression can be obtained by using Eq. (\ref{eq:displacement}) and performing a partial integration taking into account the symmetry of $\sigma_{\alpha\beta}$, \begin{equation}\label{eq:angular rate}
\dot{J}_{\alpha}=\int dA_{\delta} ~ \epsilon_{\alpha\beta\gamma}r_{\beta}\sigma_{\gamma\delta}+\frac{1}{2}\int dA_{\eta} ~ r_{\eta}\hbar\dot{S}_{\alpha}^{\prime}-\frac{1}{2}\int dA_{\alpha} ~ r_{\beta}\hbar\dot{S}_{\beta}^{\prime}.
\end{equation}
Here $dA_{\delta}$, $dA_{m}$, and $dA_{\alpha}$ refer to the boundary surfaces in the $\delta, m, \alpha=x, y, z$ directions. 

\section{numerical solution for a disk model} \label{Section:numerical solution}
A quasi-two-dimensional disk model depicted in Figure \ref{fig:mechanism} is adopted to solve Eqs. (\ref{eq:displacement}) and (\ref{eq:spin vector dynamics}). In general, the coefficient tensor $C_{\alpha\beta\gamma\delta}$ is a $9\times9$ matrix. However, for an isotropic object with symmetric stress and strain tensors, $C_{\alpha\beta\gamma\delta}$ can be described only by two independent elastic moduli $\lambda$ and $\mu$ \cite{lurie2010theory}, $C_{\alpha\beta\gamma\delta}=\lambda\delta_{\alpha\beta}\delta_{\gamma\delta}+\mu(\delta_{\alpha\gamma}\delta_{\beta\delta}+\delta_{\alpha\delta}\delta_{\beta\gamma})$. Hooke's law relates it to the stress tensor, $\sigma_{\alpha\beta}=E/(1+\nu)[u_{\alpha\beta}+\nu/(1-2\nu) u_{\gamma\gamma}\delta_{\alpha\beta}]$, where $E$ is Young's modulus given by $E={\mu(3\lambda+2\mu)}/({\lambda+\mu})$, and $\nu$ is Poisson's coefficient, defined as $\nu={\lambda}/{2(\lambda+\mu)}$. Polar coordinates are more convenient for the complanate and axisymmetric disk system (A detailed derivation of coordinate transforation is provided in Appendix \ref{supp:dynamics equations}). Assuming plain strain and solely considering $\theta$-invariant solutions, we can establish that $u_{az}=u_{za}=0$, $\nabla_{z}S_{a}=0$,  $\nabla_{\theta}u_{a}=0$ and $\nabla_{\theta}S_{a}=0$, where $a\in\{r,\theta,z\}$. 

To clearly demonstrate the influence of material properties on the evolution of the system, we substitute all variables with dimensionless counterparts, which are defined as follows,
 \begin{align}
	\overline{u}_{\theta}= \frac{u_{\theta}}{R}, \overline{r}= \frac{r}{R}, \overline{\boldsymbol{S}} = \frac{\boldsymbol{S}}{S},
	\overline{t}= tm, m\equiv \sqrt{\frac{E}{2(1+\nu)\rho R^{2}}},\label{dimensionless variables} 
 \end{align}
where $R$ is the radius of the disk. The dynamical equations for $\overline{u}_{\theta}$ (the displacement variable associated with the rotation of the disk) and $\overline{\bm{S}}$ in polar coordinates can be derived from Eqs. (\ref{eq:displacement}) and (\ref{eq:spin vector dynamics}) (For further calculation details, please refer to Appendix \ref{supp:dynamics equations}),
\begin{equation}\label{eq:lattice dynamics}
\frac{\partial ^2 \overline{u}_{\theta}}{\partial \overline{r}^2}+\frac{1}{\overline{r}}\frac{\partial \overline{u}_{\theta}}{\partial \overline{r} }-\frac{1}{\overline{r}^2}\overline{u}_{\theta}=\frac{\partial ^2 \overline{u}_{\theta}}{\partial \overline{t}^2}+K\overline{f}_{\theta},
\end{equation}
\begin{align}\label{eq:spin dynamics}
\frac{\partial \overline{\boldsymbol{S}}}{\partial \overline{t}}&=\overline{\boldsymbol{S}}\times[\alpha_1(\frac{\partial^2 }{\partial \overline{r}^2}+\frac{1}{r}\frac{\partial }{\partial \overline{r}})\overline{\boldsymbol{S}}+\frac{1}{2}\left(\frac{\partial^2 }{\partial \overline{r}\partial \overline{t}}+\frac{1}{\overline{r}}\frac{\partial }{\partial \overline{t}}\right)\overline{{u}}_{\theta}\boldsymbol{e}_{z}\nonumber\\
&\quad -\beta_1\boldsymbol{e}_{z}].
\end{align}
The coupling coefficient $K=\hbar SS_{0}\sqrt{(1+\nu)/(2E\rho R^2)}$ (with $S_{0}$ representing the constant spin density \cite{PhysRevB.79.104410}), $\alpha_1=Ia^2S/(\hbar R^2m)$, $\beta_1=g\mu_{B}B/(\hbar m)$, and $\overline{f}_{\theta}=\partial ^2 \overline{S}_{z}/(\partial \overline{t}\partial \overline{r})$. In the present paradigm, the strength of the SRC exhibits correlations with elastic coefficients ($E$ and $\nu$), density ($\rho$), and exchange interaction ($I$).

Also, based on Eq. (\ref{eq:angular rate}), the time derivative of angular momentum in polar coordinates is given,
\begin{subequations}
	\begin{align}
	\dot{J}_{r}&=0,\\
	\dot{J}_{\theta}&=\frac{1}{2}\int dA_{r} ~ r\hbar\dot{S}_{\theta},\\
	\dot{J}_{z}&=\int dA_{r} ~ r\sigma_{r\theta}+\frac{1}{2}\int dA_{r} ~ r\hbar\dot{S}_{z}.
	\end{align}
\end{subequations}
In this case, the angular momentum in the $r$-direction, $J_{r}$, is automatically conserved.

Damping is ubiquitous in most situations. To account for this effect, we introduce the first-order time derivative of $u_{\theta}$ into Eq. (\ref{eq:lattice dynamics}) and integrate the Gilbert damping \cite{10.1088/978-0-7503-1074-1} into Eq. (\ref{eq:spin dynamics}). As a result, a coupled system with damping is obtained,
\begin{gather}
\frac{\partial ^2 \overline{u}_{\theta}}{\partial \overline{r}^2}+\frac{1}{\overline{r}}\frac{\partial \overline{u}_{\theta}}{\partial \overline{r} }-\frac{1}{\overline{r}^2}\overline{u}_{\theta}=\frac{\partial ^2 \overline{u}_{\theta}}{\partial \overline{t}^2}+K\overline{f}_{\theta}+\eta\frac{\partial \overline{u}_{\theta}}{\partial \overline{t}}, \label{eq:displacement damping} \\
\dot{\overline{\boldsymbol{S}}} = \overline{\boldsymbol{S}}\times\overline{\boldsymbol{\Gamma}}+\zeta(\dot{\overline{\boldsymbol{S}}}\times\overline{\boldsymbol{S}}), \label{eq:spin damping}
\end{gather}
where
\begin{align}\label{eq:spin moment}
\overline{\boldsymbol{\Gamma}}& = \alpha_1(\frac{\partial^2 }{\partial \overline{r}^2}+\frac{1}{r}\frac{\partial }{\partial \overline{r}})\overline{\boldsymbol{S}}+\frac{1}{2}\left(\frac{\partial^2 }{\partial \overline{r}\partial \overline{t}}+\frac{1}{\overline{r}}\frac{\partial }{\partial \overline{t}}\right)\overline{{u}}_{\theta}\boldsymbol{e}_{z}\nonumber\\
&\quad -\beta_1\boldsymbol{e}_{z},
\end{align}
and $\eta$ (\textgreater 0), $\zeta$ are dimensionless damping factors. 

We combine the finite difference method \cite{5391985, abe_higashimori_kubo_fujiwara_iso_2014} and the fourth-order Runge-Kutta method \cite{Hairer1993} to numerically solve the coupled differential Eqs. (\ref{eq:displacement damping}) and (\ref{eq:spin damping}). Under specific boundary and initial conditions (to be provided later), the time evolutions of the displacement field and spin field can be determined, enabling the subsequent calculation of the mechanical angular momentum, the spin angular momentum and various energy terms.

\begin{figure*}[thb!]
\centering
\subfigure{
\begin{minipage}[b]{0.45\textwidth}
\begin{overpic}[scale=0.15]{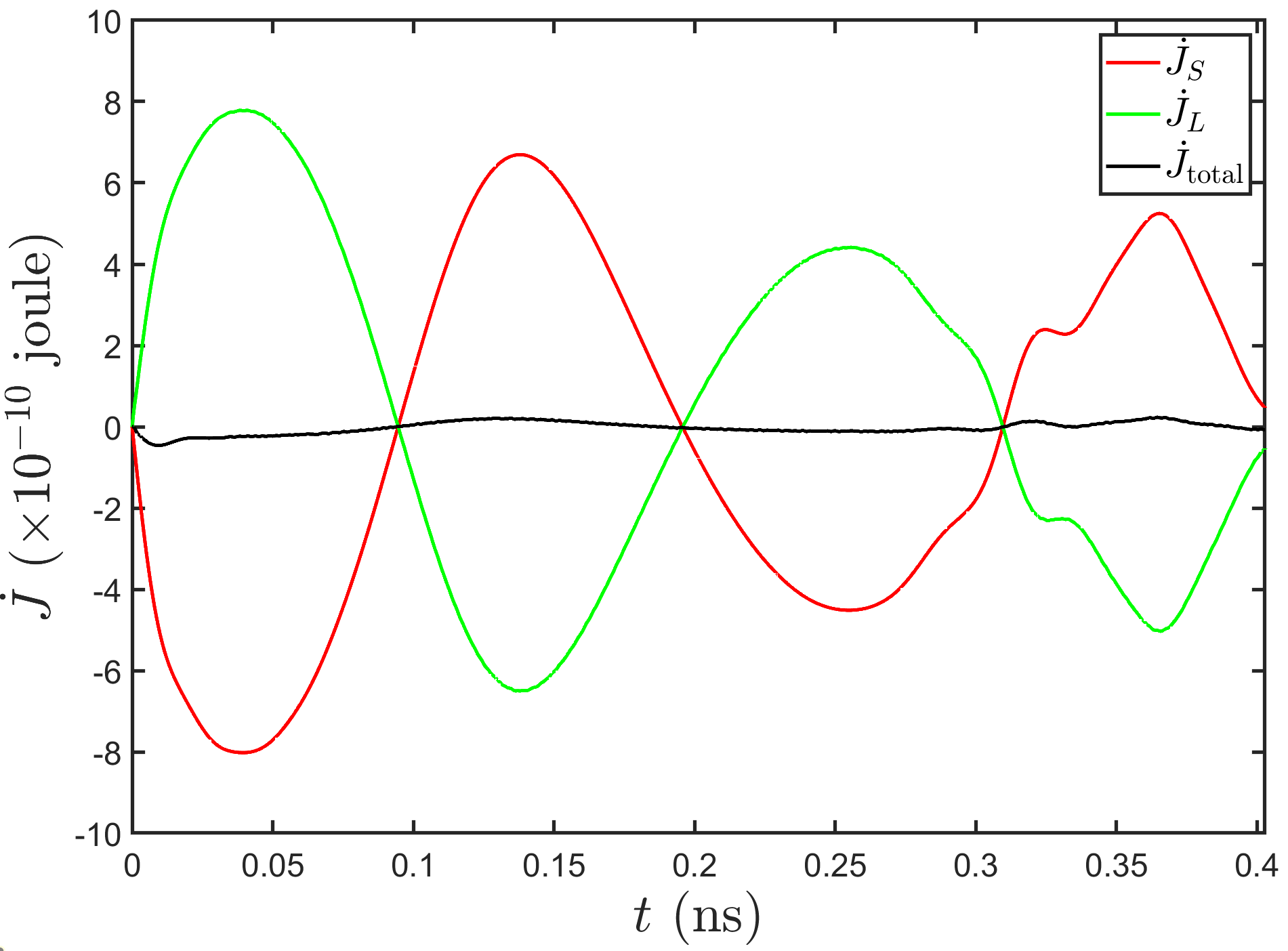}
\put(-2,68){\large\textbf{(a)}}	
\end{overpic}
\end{minipage}\label{fig:angular_nodamping}
}
\subfigure{
\begin{minipage}[b]{0.45\textwidth}
\begin{overpic}[scale=0.15]{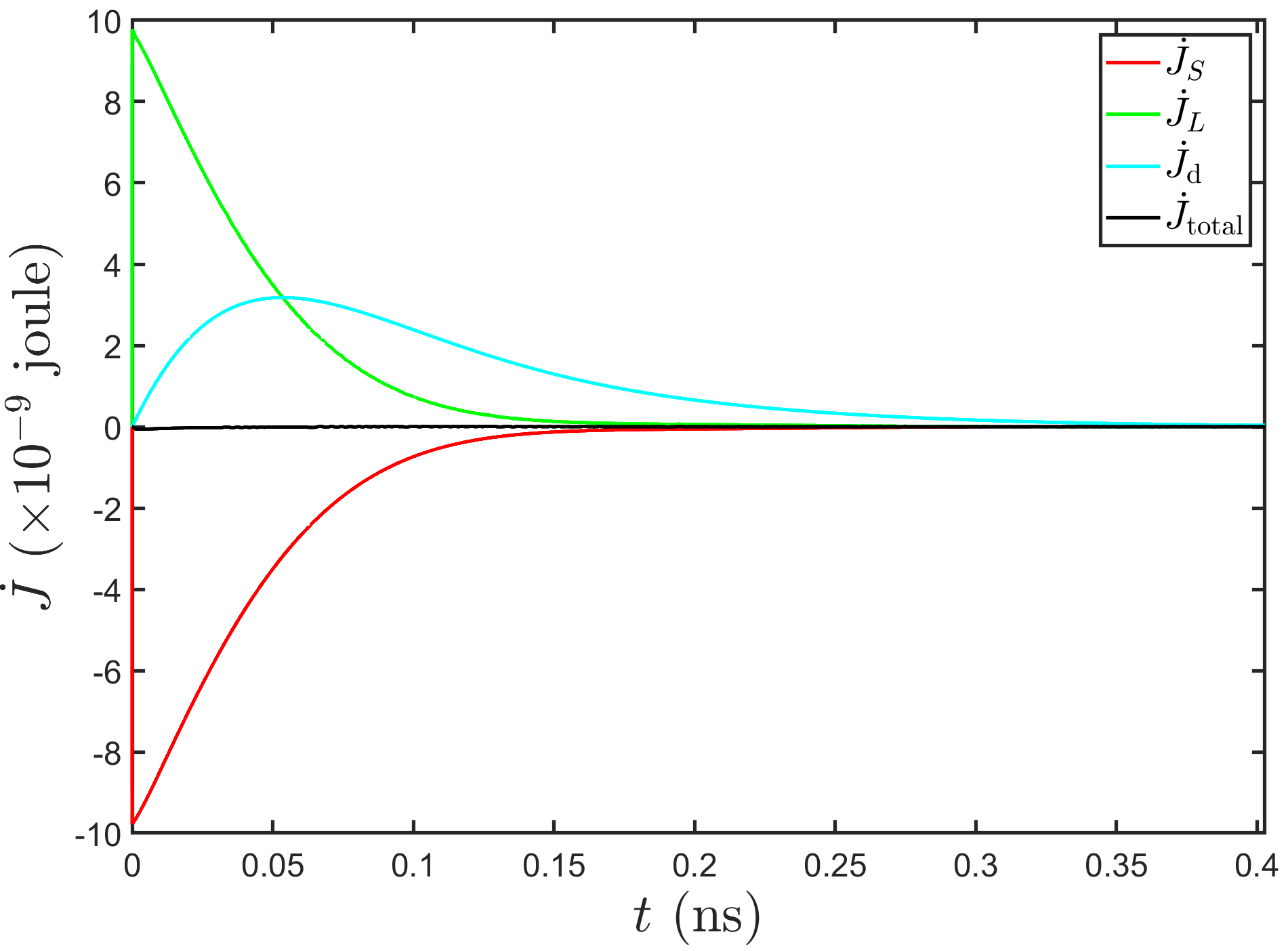}
\put(-2,68){\large\textbf{(b)}}	
\end{overpic}			
\end{minipage}\label{fig:angular_damping}
	}

\caption{Evolution of the time derivative of angular momentum in the $z$-direction. $\dot{J}_{S}$ corresponds to the spin angular momentum, $\dot{J}_{L}$ to the mechanical angular momentum, and $\dot{J}_{\mathrm{d}}$ to the angular momentum loss. (a) Undamped case, $\dot{J}_{\mathrm{totoal}}=\dot{J}_{S}+\dot{J}_{L}$. (b) Damped case, $\dot{J}_{\mathrm{total}}=\dot{J}_{S}+\dot{J}_{L}+\dot{J}_{\mathrm{d}}$. Damping factors $\eta=0.5$ and $\zeta=0.5$ are used. The large damping coefficients are chosen to display the final state of the system.}\label{fig:angular}
\end{figure*}

\section{results and analysis}\label{Section:image results}
\subsection{Transfer of angular momentum}\label{Transfer of angular momentum}
Our primary focus lies on the angular momentum in the $z$-direction, as it is relevant to the rotation of the disk. Following from the $z$-component of Eq. (\ref{eq:angular momentum derivative}), the time derivative of the spin angular momentum is
 \begin{align}\label{spin_angular}
 \dot{J}_{S}=\int d^3r ~  \hbar\dot{S}_{z},
 \end{align} 
and the time derivative of the mechanical angular momentum is
 \begin{align}\label{L_angular}
 \dot{J}_{L}=\int d^3 r ~ \rho r\ddot{u}_{\theta}.
 \end{align}
When solving Eqs. (\ref{eq:displacement damping}) and (\ref{eq:spin damping}), we specify the boundary conditions as
\begin{equation}
\left\{
\begin{aligned}
&\sigma_{r\theta}=0, ~ \sigma_{rr}=0 \qquad \mathrm{when~} r\to 0 \\
&\sigma_{r\theta}=0, ~ \sigma_{rr}=0 \qquad \mathrm{at~} r=R
\end{aligned}
\right. 
\qquad \mathrm{for~} u_{\theta}, 
\end{equation} 
and
\begin{equation}
\left\{
\begin{aligned}
&\dot{J}_{\theta}=0, ~ \dot{J}_{z}=0 \qquad \mathrm{when~} r\to 0 \\
&\dot{J}_{\theta}=0, ~ \dot{J}_{z}=0 \qquad \mathrm{at~} r=R
\end{aligned}\right.
\qquad \mathrm{for~} \boldsymbol{S}.  
\end{equation}
The selection of $r\to 0$ as a boundary instead of $r=0$ stems from the requirement to ensure the validity of the formula $1/r$. The explicit formulations of $\sigma_{r\theta}$ and $\sigma_{rr}$ can be found in Appendix \ref{supp:dynamics equations}. The first two conditions on $u_{\theta}$ correspond to the absence of force at the two boundaries, while the last two conditions on $\boldsymbol{S}$ guarantee the total angular momentum of the system cannot flow out from the boundaries. 

Since the iron disk is ferromagnetic, the following initial conditions are taken---at the beginning, all spins on the disk are uniformly aligned, and both atomic displacement and velocity are zero (i.e., the disk is stationary). It should be emphasized that our approach extends beyond the initial circumstances, and as will be seen later, we intentionally opt for other initial conditions to effectively illustrate the energy transfer process.
\subsubsection{Undamped and damped cases}\label{Damped and undamped cases}
The selected magnetic material for this study, Fe, possesses an exchange parameter of $I=4.29\times10^{-21}\mathrm{~J}$, a lattice constant of $a=2.87\times10^{-10}\mathrm{~m}$, a mass density of $\rho=7.9\times10^{3}\mathrm{~kg/m^3}$, Young's modulus of $E=1.85\times10^{11}\mathrm{~Pa}$, and Poisson's coefficient of $\nu=0.32$, as reported by \cite{AMANN2019217}. The disk radius is set to $R=10^{-7}~ \mathrm{m}$. Unless otherwise noted, these parameters are used consistently in the following calculations.

Figure \ref{fig:angular} shows the results of angular momentum transfer with and without damping. The variable $\dot{J}_d$, as shown in Fig. \ref{fig:angular_damping}, represents the rate of change of angular momentum loss attributed to the displacement damping ($\eta$). According to Eq. (\ref{eq:displacement damping}), the damping term introduces a resistive force acting on the atom, described by $F_{\mathrm{d}}^{\theta}=-\eta ~ {\partial u_{\theta}}/{\partial t}$. This leads to the emergence of a torque in the $z$-direction, which can be equivalently expressed as the time derivative of the angular momentum,
\begin{equation}\label{damping_angular}
 \dot{J}_{\mathrm{d}}=-\eta\int d^3 r ~ r\frac{\partial u_{\theta}}{\partial t}.
\end{equation}
At $t=0$, the rates of change of spin angular momentum and mechanical angular momentum are both zero as a result of the initial conditions of the system. However, at the next moment, the magnetic field $B=1\mathrm{~T}$ is applied to the disk,  leading to non-zero rates of change of angular momentum [see Eqs. (\ref{eq:displacement damping}), (\ref{eq:spin damping}), (\ref{spin_angular}), and (\ref{L_angular})]. Especially in the presence of damping, both the driving force ($\overline{f}_{\theta}$) and the first-order time derivative of the displacement field ($\dot{u}_{\theta}$) are no longer zero after the introduction of the magnetic field, causing an abrupt increase in the second-order time derivative of the displacement field ($\ddot{u}_{\theta}$), i.e., a sharp jump in the rate of change of the mechanical angular momentum in Fig. \ref{fig:angular_damping}. This phenomenon indicates that the external magnetic field acts as the driving source for the EdH effect. If the driving source is the rotation of the magnet (no external magnetic field in this case), the evolution of the displacement field will result in spin-flips, similar to the Barnett effect. Notably, between $t=0.25 ~ \mathrm{ns}$ and $t=0.35 ~ \mathrm{ns}$, there is a strange peak, which can be suppressed by a stronger magnetic field, suggesting its intrinsic cause is the competition between the external magnetic field and the intrinsic elastic dynamics. From an overall perspective, $\dot{J}_{L}$ and $\dot{J}_{S}$ exhibit a high-frequency periodic behavior in Fig. \ref{fig:angular_nodamping}. The correlation between the oscillation frequency and the strength of the external magnetic field as well as material properties will be discussed in subsequent sections.

The timescale of angular momentum transfer is another pivotal factor closely associated with spin-mechanical applications. Ref. \cite{tauchert2022polarized} reveals that when a magnetized material is impinged by an ultrafast laser, the electronic spins transfer their angular momentum to lattice atoms within a few hundred femtoseconds. This transfer leads to the generation of circularly polarized phonons, which then propagate the angular momentum across the material at the speed of sound, ultimately contributing to the macroscopic EdH rotation. Based on this research, we make a tentative estimation of the time required for the angular momentum to transfer from local rotation to macroscopic rotation, which approximates to $t\approx R/v=0.02~ \mathrm{ns}$, with the speed of sound $v\propto\sqrt{E/\rho}$. This timescale is reflected in our theoretical results. In Fig. \ref{fig:angular_nodamping}, at $t=0.02 ~ \mathrm{ns}$, the rate of change of the angular momentum is large enough to indicate the occurrence of macroscopic rotation. Considering the coefficient factor of $v$ is undetermined, we estimate the transfer timescale of spin angular momentum to the whole lattice to be on the order of 0.01 ns. Nevertheless, accurately evaluating the femtosecond timescale for the transfer of angular momentum from spin to local rotation is currently unfeasible due to the challenge of precisely identifying the exact moment when local rotation emerges. 

Throughout the coupling process, the rate of change of the spin angular momentum ($\dot{J}_{S}$) and the rate of change of the lattice angular momentum ($\dot{J}_{L}$, or the sum of $\dot{J}_{L}$ and $\dot{J}_{\mathrm{d}}$ in the damped case) evolve inversely, with their sum approximately (but not strictly) equal to zero. This situation is not improved by enhancing computational accuracy. One possible explanation for this phenomenon is the exclusion of phonon spin angular momentum from Eq. (\ref{eq:angular momentum}). The total mechanical angular momentum should incorporate the contribution of the phonon spin angular momentum \cite{PhysRevLett.112.085503, PhysRevB.103.L100409}, which is defined as $\bm{J}_{\mathrm{p-s}}=\int d^3 r~ \rho(\bm{u}_{\mathrm{local}}\times\ \dot{\bm{u}}_{\mathrm{local}})$, where $\bm{u}_{\mathrm{local}}$ represents the atomic displacement around the equilibrium position, corresponding to local rotations in Fig. \ref{fig:mechanism}. The displacement $\bm{u}$ described in Eq. 
(\ref{eq:displacement}) includes both local and overall rotations, without separate kinetic equations to describe $\bm{u}_{\mathrm{local}}$. Additionally, it is challenging to distinguish the contributions of the local displacement and overall displacement within $\bm{u}$. Therefore the calculation of $\bm{J}_{\mathrm{p-s}}$ is currently unavailable. Eq. (\ref{eq:angular momentum}) differs from the total angular momentum by $\bm{J}_{\mathrm{p-s}}$, leading to $\dot{J}_{\mathrm{total}}$ not being strictly zero, as shown in Fig. \ref{fig:angular_nodamping}. However, considering that the local displacement is much smaller than $r$, the angular momentum $J_{\mathrm{p-s,z}}$ (the $z$-component of $\bm{J}_{\mathrm{p-s}}$) is much smaller  compared to $J_{L,z}=\int d^3 r~ \rho(\bm{r}\times\ \dot{\bm{u}})_z$. This renders the contribution of it negligible in our analysis of the interconversion between spin angular momentum and lattice angular momentum within the EdH effect. 

As our approach is not constrained by initial conditions, it allows exploring the spin-lattice coupling processes where the system start from a nonequilibrium or nonferromagnetic configuration. First-principle calculations can assist in determining the initial spin configurations of diverse materials. Besides, the dynamic behavior of magnetic systems under external perturbations, including magnetic field and stress, can also be investigated. By incorporating the effects of these perturbations into the coupling equations, one can analyze the response and sensitivity of the system to different stimuli and predict their influences on the dynamical evolution from lattice to spin or from spin to lattice. 

\begin{figure}[t!]
	\centering
	\subfigure{
		\begin{minipage}[b]{0.45\textwidth}
			\begin{overpic}[scale=0.15]{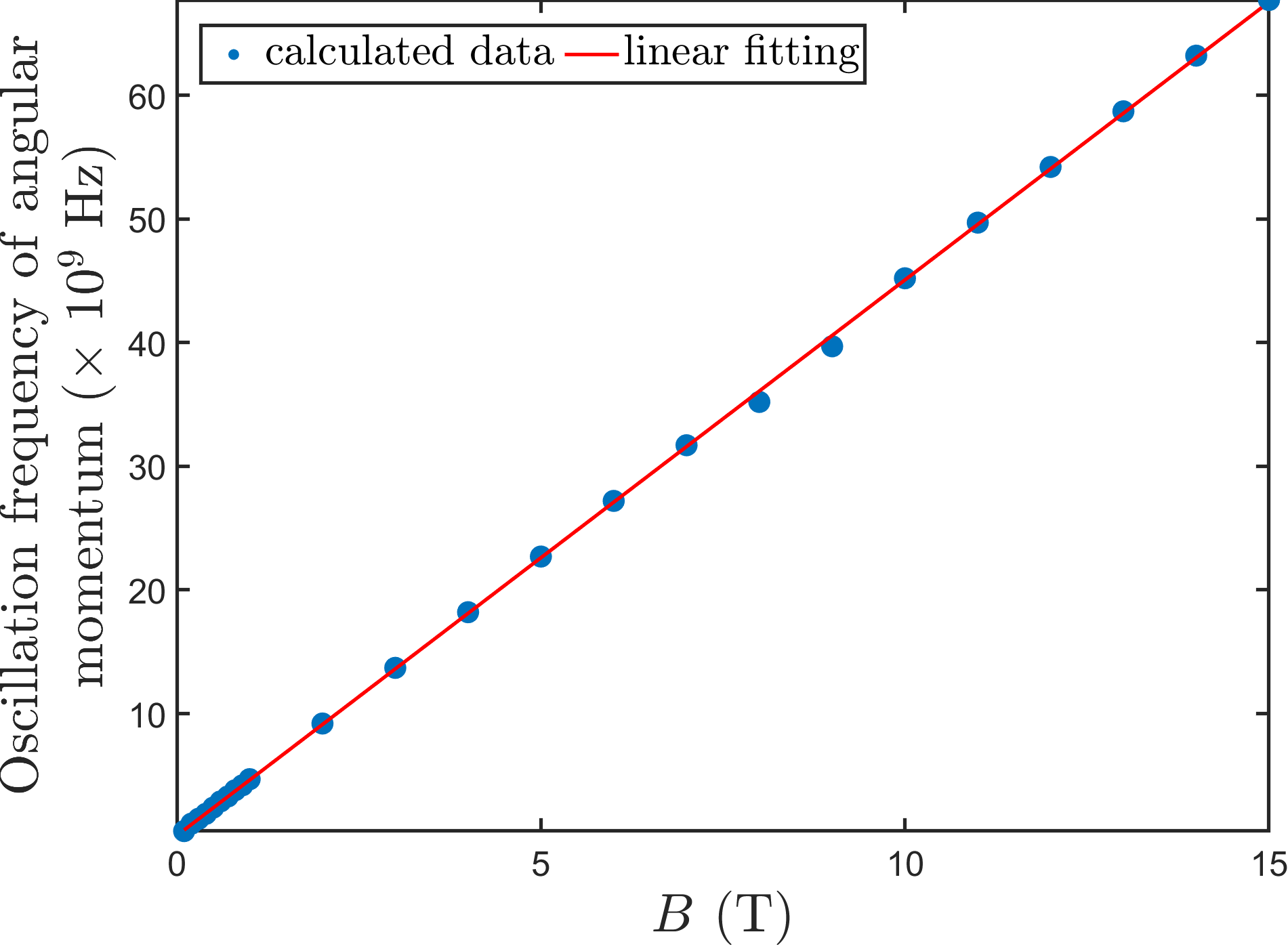}
				\put(0,73){\large\textbf{(a)}}	
			\end{overpic}
		\end{minipage}\label{fig:magnetic_frequency}
	}
	\subfigure{
		\begin{minipage}[b]{0.45\textwidth}
			\begin{overpic}[scale=0.15]{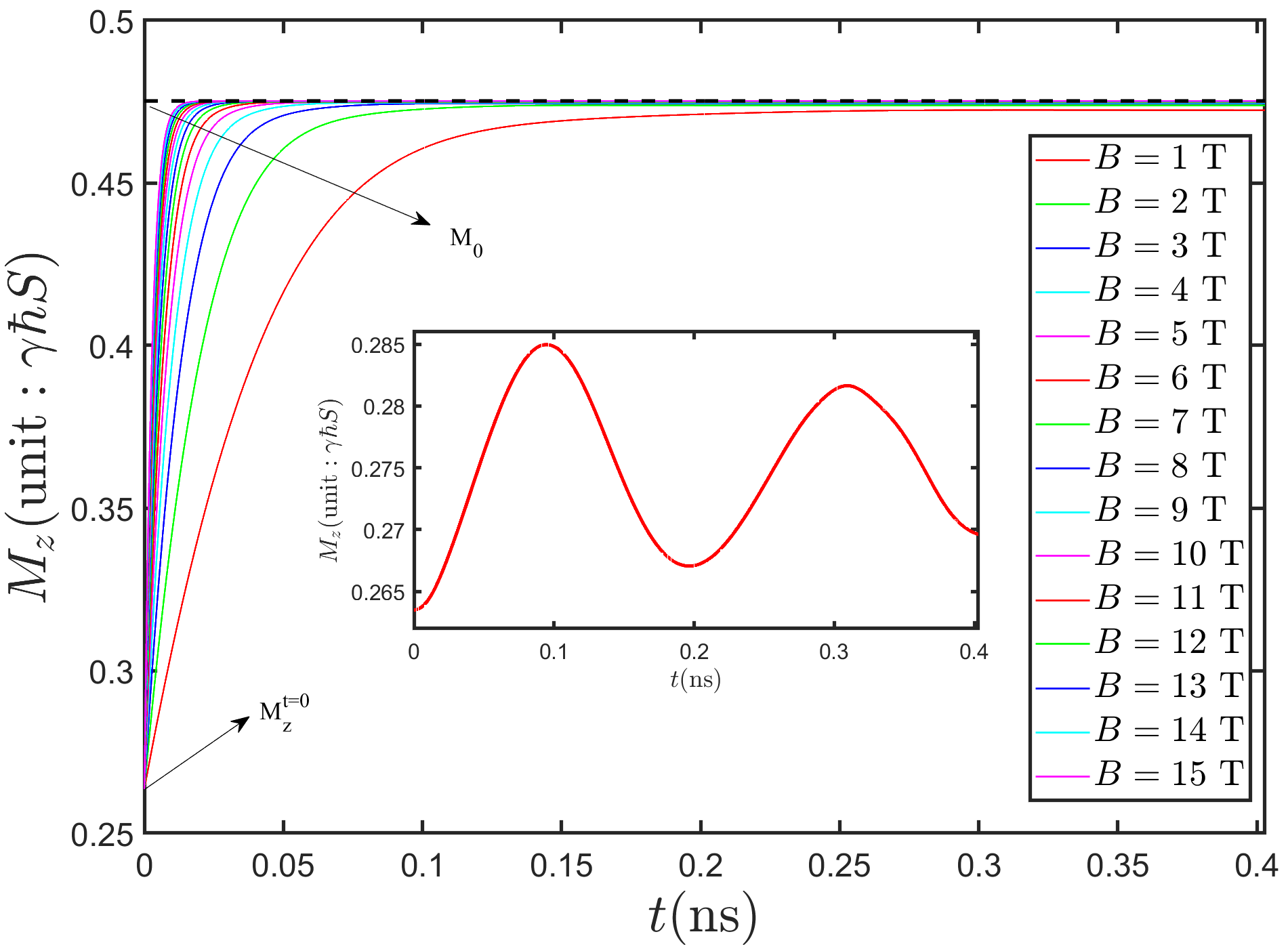}
				\put(0,75){\large\textbf{(b)}}	
			\end{overpic}			
		\end{minipage}\label{fig:M_z}
	}

\subfigure{
	\begin{minipage}[b]{0.45\textwidth}
		\begin{overpic}[scale=0.15]{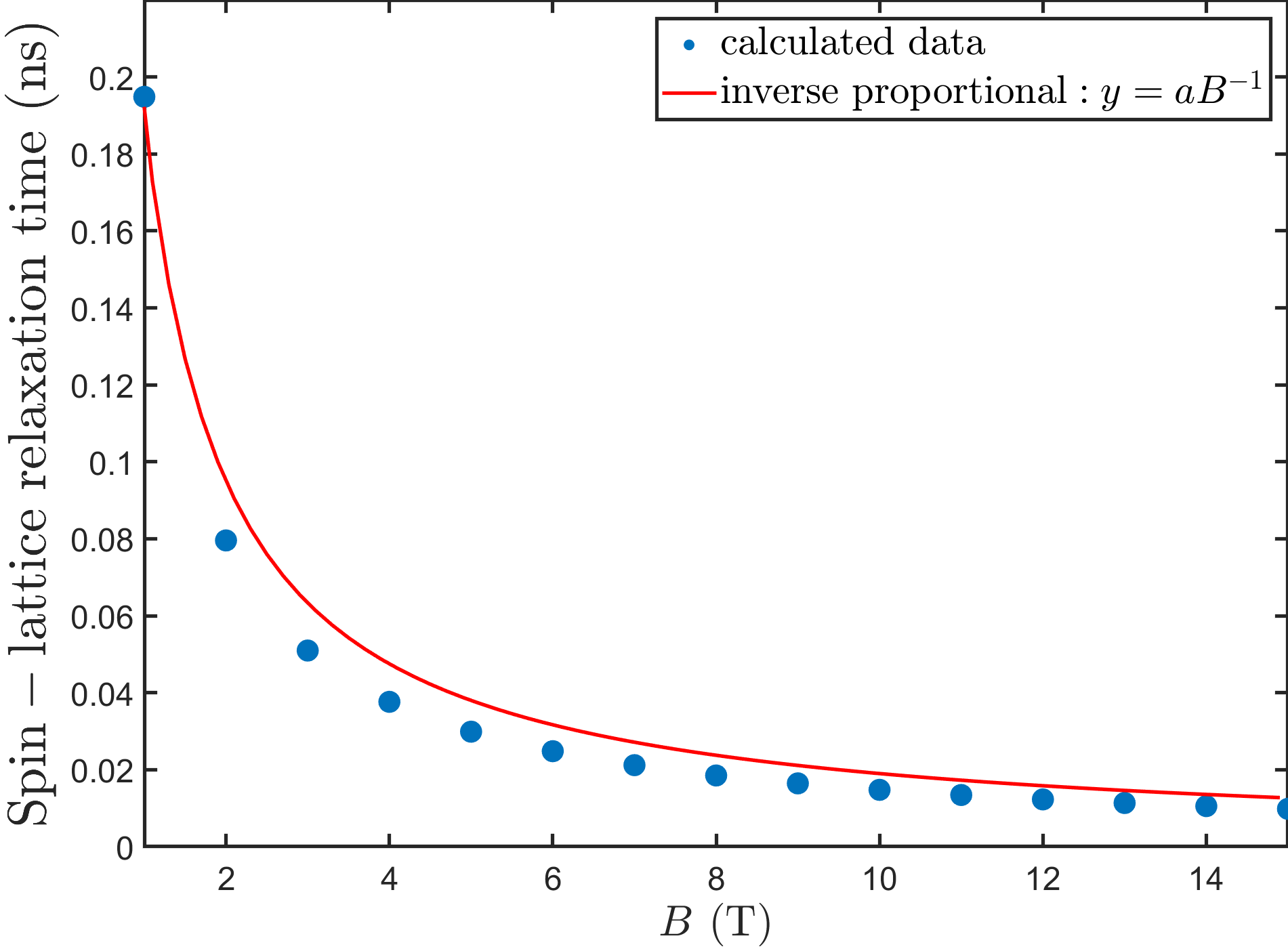}
			\put(0,75){\large\textbf{(c)}}	
		\end{overpic}			
	\end{minipage}\label{fig:magnetic_angular}
}
	\caption{Influence of the magnetic field on angular momentum transfer. (a) The oscillation frequency of angular momentum versus the magnetic field $B$. In the linear fitting $y=k_{1}B+b_{1}$, $k_{1}=4.493$ and $b_{1}=0.1367$. (b) Magnetization evolution under different magnetic fields. Damping factors $\eta=0.5$ and $\zeta=0.5$. The inset shows the case without damping, but with magnetic field $B=1~\mathrm{T}$. (c) Spin-lattice relaxation time versus $B$. The solid line (shown in red) represents an inverse proportional fitting with $a=0.19$.}\label{fig:magnetic field}
\end{figure}

\subsubsection{Effect of magnetic field}\label{effect of magnetic field}
In the following we will study the effect of a constant magnetic field on the evolution of the system. An alternating magnetic field can also be employed and without changing the main conclusions. We define the oscillation frequency of angular momentum as the number of oscillations per second, and its relationship with the magnetic field is depicted in Fig. \ref{fig:magnetic_frequency}. When there is no damping, the frequency is proportional to the magnitude of the magnetic field, $f\propto B$, and the proportionality constant is the slope of the linear fitting, $k_{1}$. Eq. (\ref{eq:displacement damping}) reveals two distinct types of displacement field evolution: intrinsic and forced. The former is governed by the purely elastic dynamics equation, and its oscillation frequency is determined by the material elasticity coefficients, while the latter is driven by the change of spin, with an oscillation frequency determined by the spin torque, $\dot{\bm{S}}=\bm{S}\times[ Ia^2\nabla^2 \bm{S}/\hbar+\gamma\bm{B}+(\nabla\times\dot{\bm{u}})/2]$. The linear relationship between the oscillation frequency and the external magnetic field suggests that the displacement field mainly follows forced oscillations. Under different magnetic fields, we compare the magnitude of spin torques generated by the Heisenberg term $Ia^2\nabla^2 \bm{S}/\hbar$, the external magnetic field $\gamma\bm{B}$, and the displacement field $(\nabla\times\dot{\bm{u}})/2$. We find that the contribution from the displacement field is significantly smaller than that from the magnetic field. Moreover, although the Heisenberg interaction torque varyies with the spin itself, it is also lower than the magnetic field torque across most of the disk. Consequently, the external magnetic field dominates the evolution frequency of the spin and displacement fields, while the other two serve as modulating factors. This reminds us that in purely precessional dynamics of spin, the precession frequency is proportional to the effective field that induces spin precession \cite{10.1088/978-0-7503-1074-1}, $\dot{\bm{S}}=\bm{S}\times\gamma\bm{H}_{\mathrm{eff}}$, $w=\lvert\gamma\bm{H}_{\mathrm{eff}}\rvert=2 \pi f^{\prime}$, where $\bm{H}_{\mathrm{eff}}$ is the effective field, $w$ is the angular frequency, and $f^{\prime}$ is the frequency. We discover that $k_{1}$ and $\lvert\gamma/(2\pi)\rvert/10^9$ are not equal, with $k_{1}$ being approximately one-sixth of the latter. However, if the spin torques solely contain the magnetic field term, the spins will precess around $\bm{B}$, and the spin angular momentum in the $z$-direction remains unchanged and is not transferred to the lattice system. In other words, despite the inferior effect on the oscillation frequency, the Heisenberg term and the displacement field term play a crucial role in facilitating the transfer of angular momentum between the spins and lattice. 

In the scenario of damping, our investigation aims at the dependence of the spin-lattice relaxation time on the magnetic field. For this purpose, we denote the $z$-component of magnetization intensity as $M_z$ and the saturation magnetization intensity as $M_0$. Figure \ref{fig:M_z} illustrates the dynamic behavior of $M_z$ under varying magnetic fields. When both the magnetic field and damping are present, all the spins end up pointing in the direction of the magnetic field, with the higher magnetic field leading to faster saturation of the magnetization. In contrast, without damping (as shown in the inset), the spins exhibit periodic oscillations and fail to align with the z-axis direction, despite the action of the magnetic field. Damping is essential in achieving a stable magnetization. The spin-lattice relaxation time is defined as the duration required for $M_z$ to reach $M_0$, which is considered achieved when the deviation is less than or equal to 0.02, i.e., $\Delta M^{z}/\Delta M\leqslant 0.02$, where $\Delta M^{z}=M_{0}-M_{z}$, $\Delta M=M_{0}-M_{z}^{t=0}$. As shown in Fig. \ref{fig:magnetic_angular}, the relaxation time decreases at a slower and slower rate as the magnetic field increases. Moreover, through data fitting, we observe an near-inverse correlation between the relaxation time and the magnetic field. This feature also appears in the Landau-Lifshitz-Gilbert (LLG) equation for the typical dissipation time $\tau_{LLG}$ \cite{PhysRevLett.95.267207}, $\tau_{LLG}=1/(w_{L}\zeta)$, where $w_{L}$ is the precession frequency and $\zeta$ is the damping factor. 

\begin{figure}[t!]
	\centering
	\subfigure{
		\includegraphics[width=0.45\textwidth]{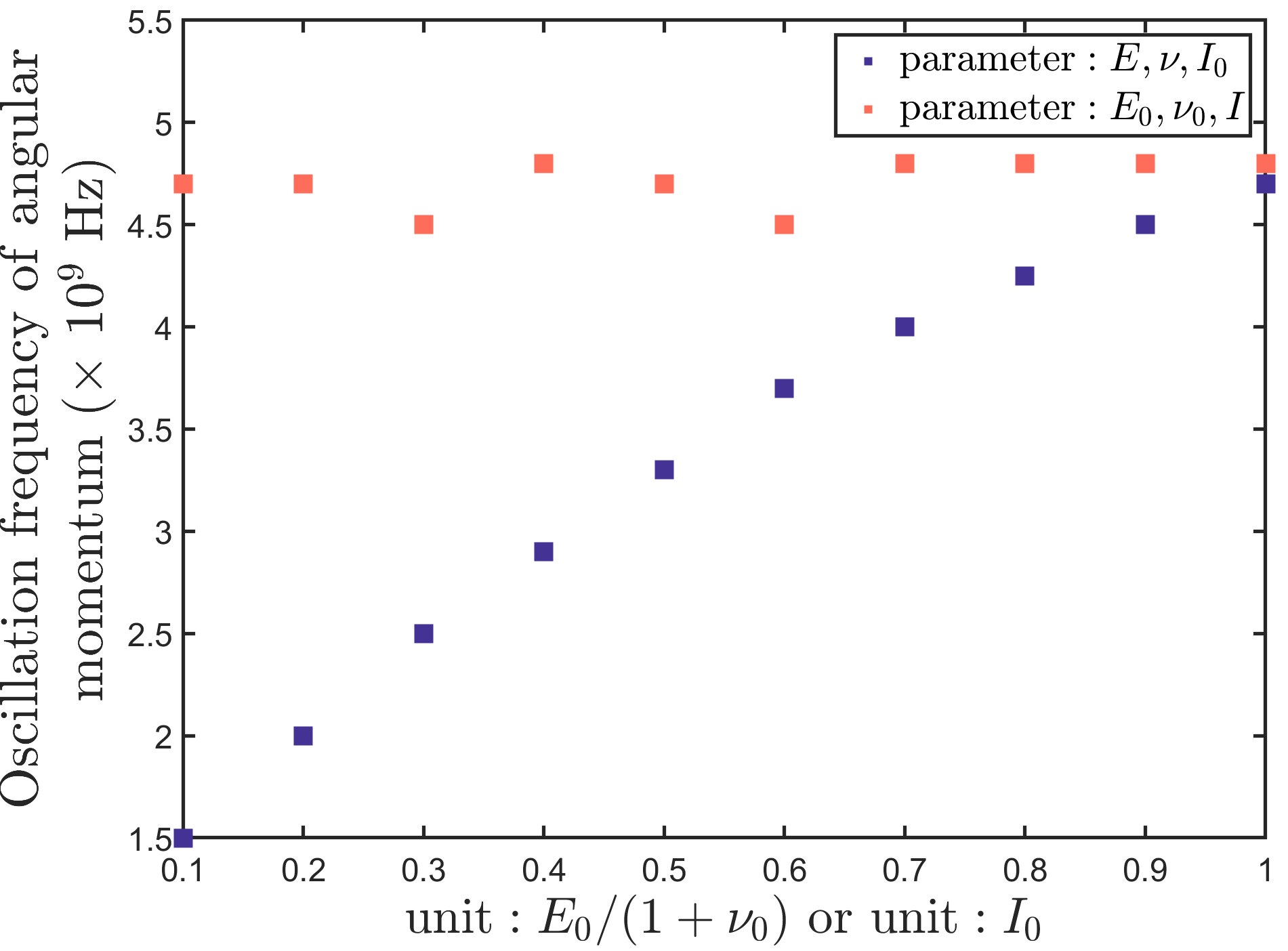}	
}
\caption{The oscillation frequency of angular momentum versus $E/(1+\nu)$ and $I$. Here, $E_{0}=1.85\times10^{11}\mathrm{~Pa}$, $\nu_{0}=0.32$, $I_{0}=4.29\times10^{-20}\mathrm{~J}$, and $B=0~\mathrm{T}$.}
	\label{fig:E_nu_angular}
\end{figure}
\begin{figure*}[thb!]
	\centering
	\subfigure{
		\begin{minipage}[b]{0.45\textwidth}
			\begin{overpic}[scale=0.15]{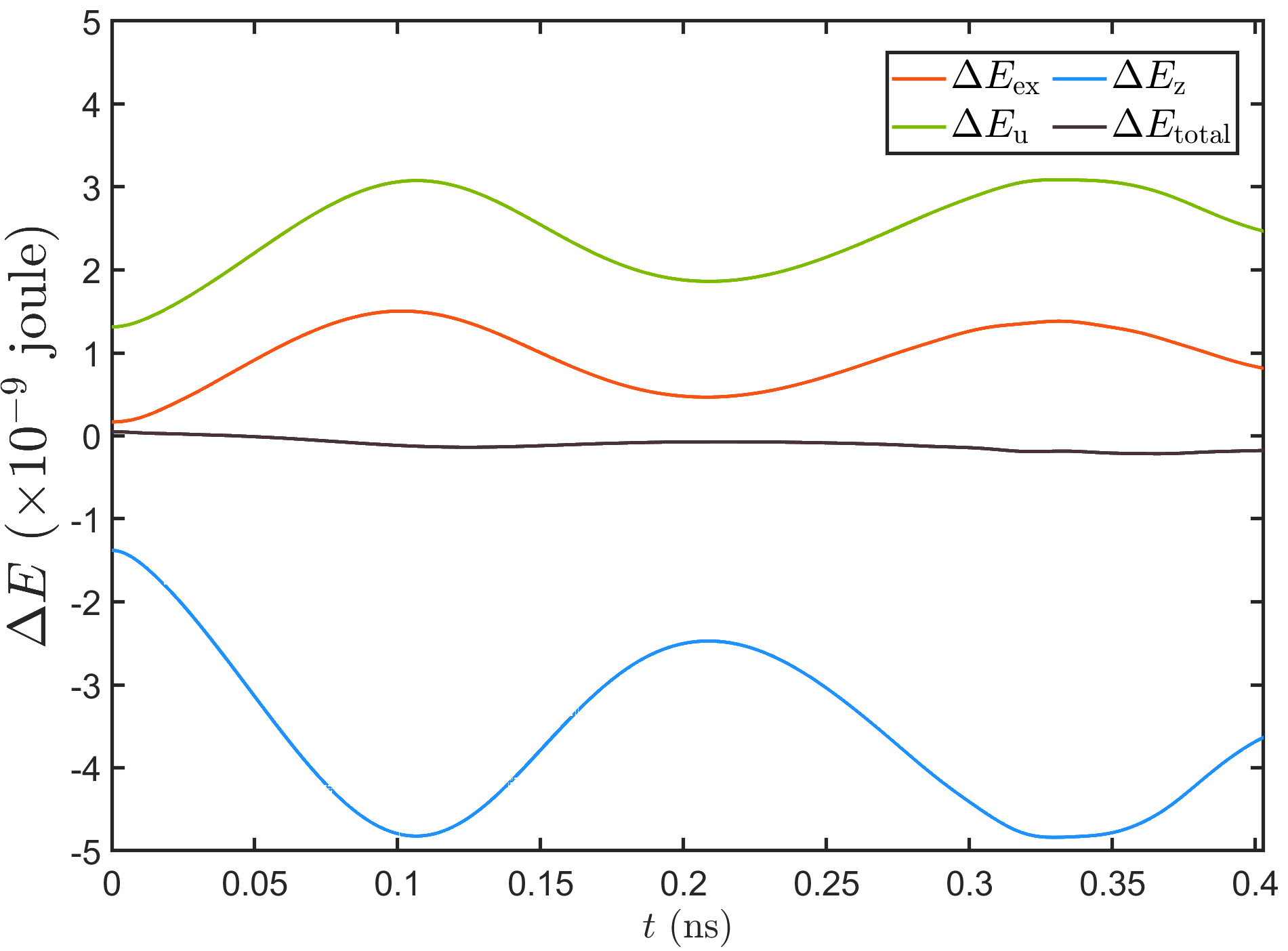}
				\put(-2,68){\large\textbf{(a)}}
			\end{overpic}
		\end{minipage}\label{fig:energy_nodamping}
	}
	\subfigure{
		\begin{minipage}[b]{0.45\textwidth}
			\begin{overpic}[scale=0.15]{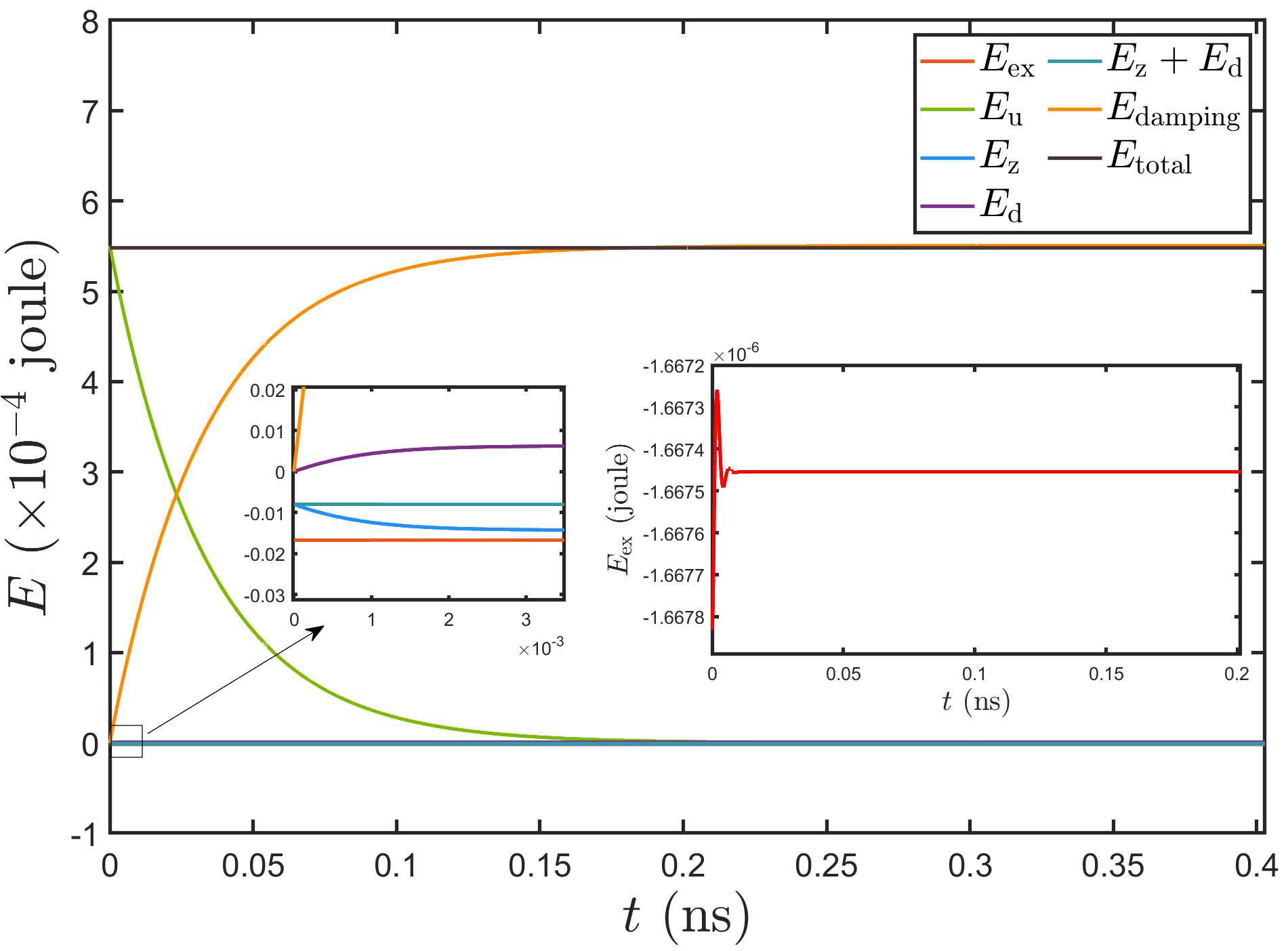}
				\put(-2,68){\large\textbf{(b)}}
			\end{overpic}			
		\end{minipage}\label{fig:energy_damping}
	}
	\caption{Evolution of each part of energy. $E_{\mathrm{ex}}$ represents the Heisenberg energy, $E_{u}$ includes the kinetic energy of the displacement field and elastic energy, $E_{z}$ is the Zeeman energy, $E_{\mathrm{d}}$ illustrates the effect of spin damping on the Zeeman term, and $E_{\mathrm{total}}$ is their sum. $\Delta$ denotes their change. (a) Undamped case with $B=2 ~ \mathrm{T}$.  (b) Damped case with $B=50 ~ \mathrm{T}$, $\zeta=0.5$, and $\eta=0.5$. $B$ is set to $50 ~ \mathrm{T}$ to amplify the variation of $E_{z}$.} \label{fig:energy}
\end{figure*}
\subsubsection{Effect of material parameters}
The variations in material parameters, such as the exchange coefficient ($I$), Young's modulus ($E$), and Poisson's coefficient ($\nu$), can also affect the systerm. But our calculations demonstrate that when the system is exposed to a magnetic field, these variations can significantly impact the magnitude of the angular momentum while having minimal effect on the frequency. Therefore, we exclude the magnetic field ($B=0$) and utilize the rotation of the disk as the driving force to explore the influence of these material parameters on the evolution frequency. The chosen initial angular velocity is the same as that used in the energy transfer case discussed below. The oscillation frequency of angular momentum in Fig. \ref{fig:E_nu_angular} ranges from $4.5\times10^{9}$ Hz to $5\times10^{9}$ Hz, with negligible dependence on $I$. This behavior is expected since $I$ does not directly affect the coupling coefficient $K$ for isotropic objects. However, for the elasticity coefficients $E$ and $\nu$, the frequency exhibits a positive correlation with the ratio of $E$ to $\nu$, not in a linear manner like the magnetic field. The Young's modulus of a material is a manifestation of its interatomic bonding energy and lattice structure. A larger $E$ to $\nu$ ratio indicates a harder material, which is more likely to move as a whole and less prone to local elastic deformation. This information is useful for selecting suitable elastic materials as potential candidates for magneto-mechanical devices.

\subsection{Transfer of energy}\label{Transfer of energy}
The EdH process also involves an energy exchange between the spin subsystem and lattice subsystem. Noether's theorem determines the total energy as
\begin{align}\label{eq:totoal_energy}
H&=\int d^3 r ~ \left[\frac{\partial \mathcal{L}}{\partial \dot{\Phi_{a}}}\Phi_{a}-\mathcal{L}\right]\nonumber \\& =\int d^3 r ~ \bigg[\frac{1}{2}\rho\dot{\boldsymbol{u}}^{2}-\frac{I^{\prime}}{2}\boldsymbol{S}^{\prime}\cdot(\boldsymbol{S}^{\prime}+a^2\nabla^2\boldsymbol{S}^{\prime})+g\mu_{B}\boldsymbol{B}\cdot\boldsymbol{S}^{\prime}\nonumber \\&\quad +\frac{1}{2}C_{\alpha\beta\gamma\rho}u_{\gamma\rho}u_{\alpha\beta}\bigg],
\end{align}
with $\Phi_{a}=u_{a}$,$S^{\prime}_{a}$. If the initial conditions mentioned in Section \ref{Transfer of angular momentum} persist, the lattice energy will start at a level lower than the spin energy due to the absence of lattice kinetic energy and the non-zero spin energy of the ferromagnetic ground state. To balance the consideration of spin and lattice, we adopt an initial state where the angular velocity disk is set at one $m ~ \mathrm{rad/s}$, with $m$ being the dimensionless factor of time in Eq. (\ref{dimensionless variables}).

Figure \ref{fig:energy} describes the energy conversion with and without damping. In the undamped scenario, we are concerned with the changes in energy of each component, denoted as $\Delta E$, which are all much smaller than the components themselves. As depicted, the Zeeman term $\Delta E_{\mathrm{Z}}$ evolves opposite to both the exchange term $\Delta E_{\mathrm{ex}}$ and the lattice term $\Delta E_{\mathrm{u}}$, representing the transfer of energy from the former to the latter two. Similar to angular momentum, $\Delta E_{\mathrm{total}}$ is not constant at zero. Eq. (\ref{eq:totoal_energy}) represents the conserved total energy derived from Noether’s theorem, where $\bm{u}$ is the displacement vector of the atom. However, as we only consider $u_{\theta}$ in relation to the EdH rotation, the kinetic energy $\rho{\dot{u}}^2_{\theta}$ term deviates from $\rho{\dot{\bm{u}}}^2$, making the calculated $E_{\mathrm{total}}$ not strictly constant in Fig. \ref{fig:energy_nodamping}. It is worth emphasizing that our focus is more on comprehending the mechanisms and patterns of angular momentum and energy transfers between the spin and lattice systems than the strict conservation of these quantities. In Figs. \ref{fig:angular_nodamping} and \ref{fig:energy_nodamping}, the evolution of angular momentum and energy within the spin subsystem exhibits an opposite trend compared to that of the lattice subsystem. This observation strongly indicates that the coupled Eqs. (\ref{eq:displacement}) and (\ref{eq:spin vector dynamics}) effectively capture the transfer of angular momentum and energy.

Given that the Heisenberg energy $E_\mathrm{ex}$ defined in Eq. (\ref{eq:Lagrangian density}) is directly proportional to  $\boldsymbol{S}\cdot\nabla^2\boldsymbol{S}$, we have $(\partial {E}_\mathrm{ex}/{\partial t})\propto\dot{ \boldsymbol{S}}\cdot\nabla^2\boldsymbol{S}$, where the term $\nabla^2\dot{\bm{S}}$ is neglected as it is small compared to $\dot{\bm{S}}$. With this in mind, the energy dissipation arising from the magnetic field response to spin damping, $E_{\mathrm{d}}$, satisfies $(\partial E_{d}/\partial t)\propto -\dot{\bm{S}}_\mathrm{damping}\cdot g\mu_{B}\bm{B}$. Therefore, the energy loss can be estimated as
\begin{equation}
E_{d}=-\int dt ~ g\mu_{B}\bm{B}\cdot[\frac{\zeta}{\hbar S}(\boldsymbol{S}\times\boldsymbol{\Gamma})\times\boldsymbol{S}].
\end{equation}
The energy losses resulting from damping effects are discussed in Appendix \ref{loss of angular momentum and energy}, and their total is represented as $E_\mathrm{damping}$ in Fig. \ref{fig:energy_damping}. Under large damping conditions, the kinetic energy $E_{\mathrm{u}}$ rapidly diminishes, eventually reaching zero. It is clear that $E_{\mathrm{u}}$ is the primary contributor to energy loss, combined with the trend of the loss term $E_\mathrm{damping}$. The Zeeman energy is nearly balanced out by the damping term $E_{\mathrm{d}}$, as evidenced by their constant sum shown in the inset. The change in exchange energy $E_{\mathrm{ex}}$ is relatively insignificant compared to the other energy terms, thus its behavior is shown separately in the inset. Initially, $E_{\mathrm{ex}}$ rises sharply, then falls rapidly and stabilize. It should be noted that the damping factors used in the theoretical calculations are higher than the actual values, necessitating the use of realistic damping factors to explore the specific transfer of each energy term.

\begin{figure}[t!]
	\centering
	\includegraphics[width=0.45\textwidth]{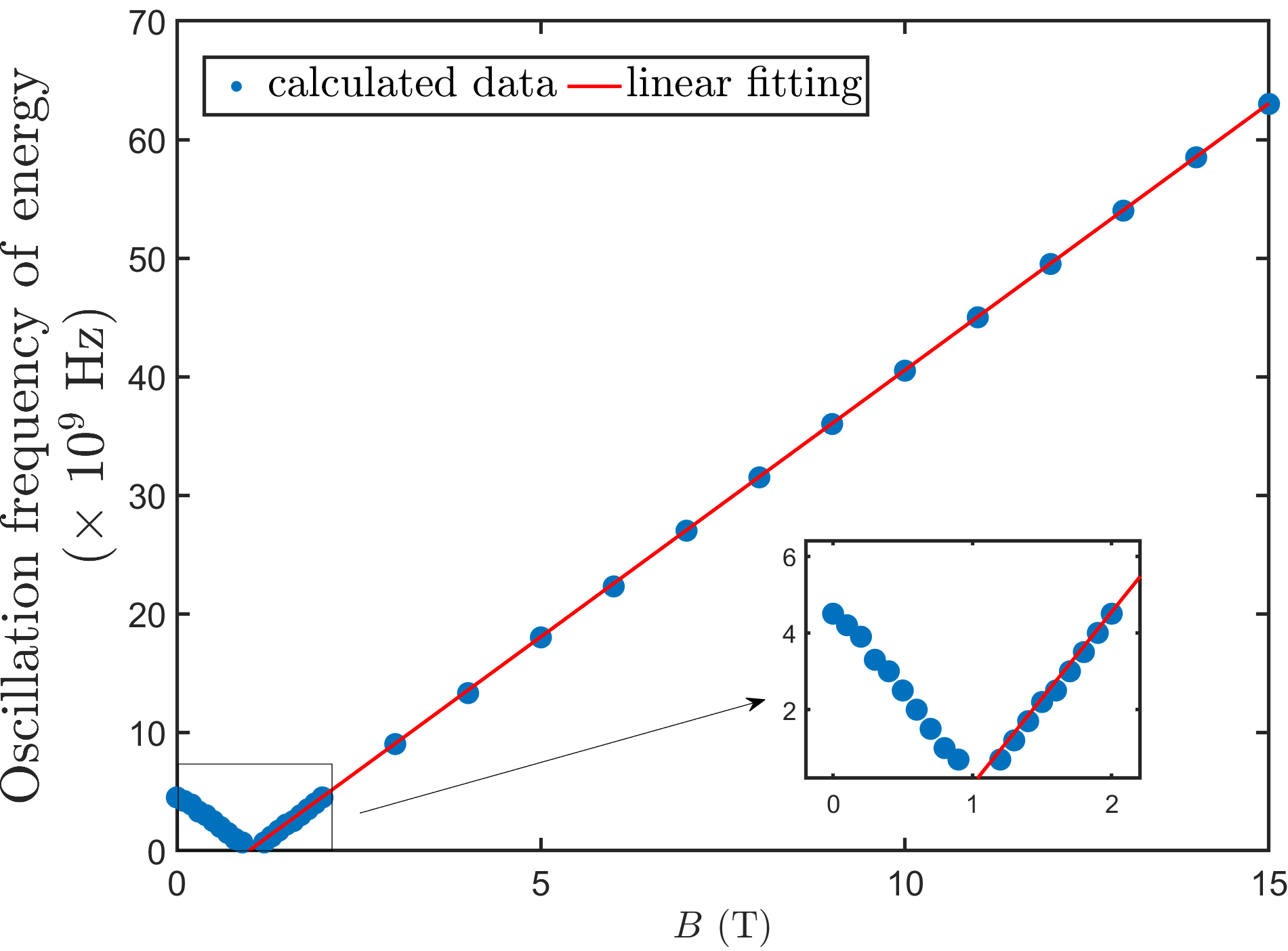}
	\caption{The oscillation frequency of energy versus the magnetic field $B$. In the linear fitting $y=k_{2}B+b_{2}$, $k_{2}=4.5$ and $b_{2}=-4.45$. The critical magnetic field (corresponding to the field at zero frequency) is about $1~\mathrm{T}$. }
	\label{fig:magnetic_energy}
\end{figure}

The effect of the magnetic field on the oscillating frequency of energy is illustrated in Fig. \ref{fig:magnetic_energy}. When the magnetic field is weak, the oscillation frequency decreases as the magnetic field strength increases. However, once the magnetic field reaches approximately $1~\mathrm{T}$, the frequency starts to linearly increase with the magnetic field. In energy calculations, we consider a non-zero initial velocity for the disk, which drives the intrinsic evolution of the system. At relatively low magnetic fields, the initial rotation and the magnetic field jointly determine the frequency of the system, with the former contributing more. But, as the magnetic field continues to grow to about $1~\mathrm{T}$, the frequency approaches 0, revealing a competitive relationship between these two driving sources in determining the frequency. With further increase in the magnetic field, the field will surpasse the influence of the initial conditions and become the dominant factor in determining the frequency. Different magnetic materials and initial conditions result in distinct critical magnetic fields. In addition, the slope of this linear correlation differs from that of the zero initial velocity case depicted in Fig. \ref{fig:magnetic_frequency}. In combination with the discussion in Section \ref{effect of magnetic field}, it can be suggested that the linear slope between the frequency and magnetic field is determined by the Heisenberg interaction, the spin-lattice coupling term, and the intrinsic evolution of the system.

\subsection{Barnett effect}
As the inverse effect of EdH, the Barnett effect is another manifestation of SRC. In this context, the driving source of the system is attributed to the rotation of the disk, not the magnetic field. Consequently, the dynamic equations take the following form,
\begin{gather}
\rho\frac{\partial^2 \bm{u}}{\partial t^2}-\nabla\cdot\bm{\sigma}+\frac{1}{2}\nabla\times\hbar\dot{\bm{S}^{\prime}}=0,\\
\dot{\bm{S}}=\bm{S}\times [Ia^2/(\hbar\nabla^2 \bm{S})+\bm{\Omega}],\label{eq:barnett}
\end{gather}
with angular velocity $\bm{\Omega}=(\nabla\times\dot{\bm{u}})/2$. Eq. (\ref{eq:barnett}) indicates that the magnetization generated by rotating a ferromagnetic object at $\bm{\Omega}$ in the absence of a magnetic field is equivalent to that produced when the object is not rotating but subjected to a magnetic field $\bm{B}_{\mathrm{eff}}= \bm{\Omega}/\gamma$. This demonstrates the Barnett effect, and the effective field $\bm{B}_{\mathrm{eff}}$ is referred to as the Barnett field. 
The measurement of the Barnett effect is commonly performed using the magnetic resonance technique, which can detect the frequency shift signal $\bm{\Omega}/\gamma$ originating from the Barnett field \cite{Chudo_2014, doi:10.7566/JPSJ.84.043601}. The study of the Barnett effect has developed rapidly, including the observation of the electronic Barnett effect in paramagnetic states \cite{PhysRevB.92.174424} and the first reported observation of the nuclear Barnett effect \cite{PhysRevLett.122.177202}. Considering that the paramagnetic and ferromagnetic states are distinguished by the Curie temperature, we can extend the existing spin-rotation theory by incorporating temperature modulation based on the Landau phase transition theory. This approach enables us to explore the spin-lattice dynamics behavior of the paramagnetic state when high-temperature conditions are taken. However, the transfer mechanism of angular momentum within nuclear systems is significantly complicated by the presence of strong interactions and electromagnetic interactions, although the principle of angular momentum conservation remains valid at any level. A possible approach to exploring the Barnett effect in nuclear systems is through the equation of motion for nuclear spin, which incorporates nuclear spin-orbit interactions. Of course, the specific issue falls beyond the scope of this paper, and requires further in-depth research in conjunction with knowledge of atomic nuclear physics \cite{article, XIA2020135130, article, universe9010036}. Besides, Ref. \cite{takahashi2016spin} has experimentally demonstrated the coupling between the mechanical rotation of the liquid metal mercury and the electron spin. The form of the spin-rotation coupling term is consistent with that of the Barnett effect in solids, which can be regarded as the Barnett effect in liquids. During this process, the angular momentum is transferred from the fluid to the spin, resulting in the generation of a spin current \cite{PhysRevB.96.020401}.

\section{CONCLUSIONS}\label{Section:conclusions}
In summary, we have successfully employed the SRC mechanism to establish a spin-lattice dynamical system capable of demonstrating the transfers of angular momentum and energy associated with the EdH effect and the Barnett effect. Through the utilization of classical field theory, our proposed approach exhibits remarkable versatility, allowing for its application to various elastic materials and accommodation of different initial conditions, including both damped and undamped scenarios.

Our calculations reveal that the transfer of angular momentum from spins to the entire lattice occurs on a timescale of approximately 0.01 ns for a disk with a radius of 100 nm. Furthermore, we observe that the evolution frequency of the system exhibits a linear dependence on the strength of the magnetic field, with the slope determined by the Heisenberg interaction, the spin-lattice coupling, and the initial state. In the presence of damping, the spin-relaxation time shows an inverse relationship with the magnetic field, resembling the typical dissipation time described in the LLG equation. Additionally, when the rotation of the disk and the external magnetic field are simultaneously used as the driving source of the system, the two effects compete with each other, and the critical magnetic field is determined by the intrinsic evolution of the system. We also find that increasing the ratio of Young's modulus to Poisson's coefficient can quantitatively raise the frequency, while the exchange interaction has no impact on it.  Recently, density functional theory has been applied to  investigate the influence of displacement field or strain on the spin exchange interaction \cite{li2021nature,PhysRevB.99.104302,PhysRevB.105.104418,PhysRevB.91.100405}, which can be taken into account in our approach to make the calculations more realistic. In addition, the incorporation of phonon spin angular momentum into our system would be valuable to further complement the microscopic mechanisms behind the ultrafast demagnetization phenomenon.

The present work offers an intuitive and fundamental study on the spin-rotation effect, which has the potential to inspire further insights and ideas. For instance, the spin-rotation Hamiltonian $H_\mathrm{s-r}$ in Eq. (\ref{eq:Coupling}) currently considers only low-order coupling, while higher-order effects, including but not limited to those from the Zeeman term, exchange term \cite{PhysRevB.79.104410}, need to be further investigated. In this regard, we suggest that the vibration of the disk in the $z$-direction should be duly considered to conserve the three angular momentums in the $r$, $\theta$, and $z$ directions. The additional spin-lattice mechanisms are also allowed by our approach, as long as they can be characterized by the spin field, displacement field, stress, or strain. On the other hand, the existence of the EdH effect in topological magnon insulator \cite{PhysRevResearch.3.023248} and the mechanical spin currents \cite{PhysRevLett.106.076601, PhysRevB.87.180402, IEDA201452, 10.3389/fphy.2015.00054} generated by the spin-rotation effect have been predicted theoretically, and both are currently active research areas.

\section*{Acknowledgements}
We thank Kun Cao, Muwei Wu, and Shanbo Chow for the helpful discussions. X. N., J. L., and D. X. Y. are supported by NKRDPC-2022YFA1402802, NKRDPC-2018YFA0306001, NSFC-92165204, NSFC-11974432, and Leading Talent Program of Guangdong Special Projects (201626003). T.D. acknowledges hospitality of KITP. Apart of this research was completed at KITP and was supported in part by the National Science Foundation under Grant No. NSFPHY-1748958.

\bibliography{aapmsamp}

\onecolumngrid

\clearpage

\setcounter{figure}{0}
\setcounter{section}{0}
\setcounter{equation}{0}
\renewcommand{\theequation}{S\arabic{equation}}
\renewcommand{\thefigure}{S\arabic{figure}}

\onecolumngrid
\appendix

\section{Elastic dynamics equations in polar coordinates} \label{supp:dynamics equations}
In polar coordinates, we have the transformation matrix
\begin{align}
\mathcal{R}(\theta)=\left(
\begin{array}{cccccc}
\cos{\theta} & \sin{\theta}\\
-\sin{\theta} & \cos{\theta}
\end{array}\right),
\end{align}
and
\begin{align}
\left(
\begin{array}{cccccc}
\nabla_{r}\\
r^{-1}\nabla_{\theta}
\end{array}\right)
=\mathcal{R}(\theta)\left(
\begin{array}{cccccc}
\nabla_{x}\\
\nabla_{y}
\end{array}\right), ~
\left(
\begin{array}{cccccc}
\bm{e}_{r}\\
\bm{e}_{\theta}
\end{array}\right)
=\mathcal{R}(\theta)\left(
\begin{array}{cccccc}
\bm{e}_{x}\\
\bm{e}_{y}
\end{array}\right), ~
\left(
\begin{array}{cccccc}
u_{r}\\
u_{\theta}
\end{array}\right)
=\mathcal{R}(\theta)\left(
\begin{array}{cccccc}
u_{x}\\
u_{y}
\end{array}\right).
\end{align}
The stain tensor in this coordinate system is just the rotated result using $\mathcal{R}(\theta)$,
\begin{align}
\left(
\begin{array}{cccccc}
\tilde{u}_{rr} & \tilde{u}_{\theta r}\\
\tilde{u}_{r\theta} & \tilde{u}_{\theta\theta}
\end{array}\right)
&=\mathcal{R}(\theta)\left(
\begin{array}{cccccc}
u_{xx} & u_{xy}\\
u_{yx} & u_{yy}
\end{array}\right)\mathcal{R}^{T}(\theta)\nonumber\\
&=\mathcal{R}(\theta)\left[\left(
\begin{array}{cccccc}
\nabla_{x}\\
\nabla_{y}
\end{array}\right)
\left(u_{x}, u_{y}\right)\right]\mathcal{R}^{T}(\theta)\nonumber\\
&=\left(
\begin{array}{cccccc}
\nabla_{r}u_{r} & \nabla_{r}u_{\theta}\\
(\nabla_{\theta}u_{r}-u_{\theta})/r & (\nabla_{\theta}u_{\theta}+u_{r})/r
\end{array}\right).
\end{align}
In general, $\tilde{u}_{\theta r}\neq \tilde{u}_{r\theta}$, because the rotation generated by the coordinate transformation is a field rather than a uniform rotation everywhere, thus the following symmetric strain tensor $u$ is defined as
\begin{equation}
u_{ab}=(\tilde{u}_{ab}+\tilde{u}_{ba})/2,
\end{equation}
where $a$, $b$ represent the direction indexes. Explicitly,
\begin{align}
u_{rr}
&=\frac{\partial u_{r}}{\partial r},\\
u_{\theta\theta}
&=\frac{u_{r}}{r}+\frac{1}{r}\frac{\partial u_{\theta}}{\partial \theta},\\
u_{r\theta}=u_{\theta r}
&=\frac{1}{2}\left(\frac{1}{r}\frac{\partial u_{r}}{\partial \theta}+\frac{\partial u_{\theta}}{\partial r}-\frac{u_{\theta}}{r}\right).
\end{align}
Under the assumption of plain strain conditions and considering only $\theta$-invariant solutions, the above expressions hold $u_{az}=u_{za}=0$, $\nabla_{\theta}u_{a}=0$, where $a\in\{r,\theta,z\}$.
Since $C$ is rotationally invariant, we still have $\sigma=Cu$. i.e.
\begin{align}
\sigma_{rr}
&=\frac{E}{(1-\nu^2)}\left[\frac{\partial u_{r}}{\partial r}+\nu\frac{u_{r}}{r}\right]\label{cylindrical strain 1},\\
\sigma_{\theta\theta}
&=\frac{E}{(1-\nu^2)}\left[\nu\frac{\partial u_{r}}{\partial r}+\frac{u_{r}}{r}\right],\\
\sigma_{r\theta}
&=\frac{E}{2(1+\nu)}\left(\frac{\partial u_{\theta}}{\partial r}-\frac{u_{\theta}}{r}\right)\label{eq:s10}.
\end{align}
Therefore, the equations of motion in polar coordinates are
\begin{align}
\rho\frac{\partial ^{2} u_{a}}{\partial t^{2}}
-f^{(R)}_{a}=&\mathcal{R}_{a\alpha}\nabla_{\beta}\sigma^{(M)}_{\alpha\beta}\nonumber\\
=& \nabla_{\beta}\sigma^{(M)}_{a\beta}
-(\nabla_{\beta}\mathcal{R}_{a\alpha})\sigma^{(M)}_{\alpha\beta}\nonumber\\
=& (\mathcal{R}^T)_{\beta b}\nabla_{b}(\sigma^{(M)}_{ac}\mathcal{R}_{c \beta})
-(\mathcal{R}^T)_{\beta b}(\nabla_{b}\mathcal{R}_{a\alpha})\sigma^{(M)}_{\alpha c}\mathcal{R}_{c\beta}\nonumber\\
=& \nabla_{b}\sigma^{(M)}_{ab}
+(\mathcal{R}^T)_{\alpha b}(\nabla_{b}\mathcal{R}_{c\alpha})\sigma_{ac}^{(M)}-(\mathcal{R}^T)_{\alpha c}(\nabla_{b}\mathcal{R}_{a\alpha})\sigma^{(M)}_{cb},
\end{align}
here $a$ and $b$, $\alpha$ and $\beta$ refer to the direction indexes in the polar and Cartesian coordinate systems, respectively. Note that $\nabla_a \mathcal{R}_{b\alpha}$ is non-zero only when $a=\theta$, then
\begin{align}
(\mathcal{R}^T)_{\alpha b}(\nabla_{b}\mathcal{R}_{c\alpha})\sigma_{ac}^{(M)}=& r^{-1}\sigma^{(M)}_{ac}(\nabla_{\theta}\mathcal{R}_{c\alpha})(\mathcal{R}^T)_{\alpha \theta}\nonumber\\
=& r^{-1}\sigma^{(M)}_{ac}
\begin{pmatrix}
-\sin\theta & \cos\theta\\
-\cos\theta & -\sin\theta\\
\end{pmatrix}_{c\alpha}
\begin{pmatrix}
-\sin\theta\\
\cos\theta
\end{pmatrix}_{\alpha\theta}\nonumber \\
=& r^{-1}\begin{pmatrix}
\sigma_{ar}^{(M)} \\
0
\end{pmatrix},
\end{align}
\begin{align}
-(\mathcal{R}^T)_{\alpha c}(\nabla_{b}\mathcal{R}_{a\alpha})\sigma^{(M)}_{cb}
=&-r^{-1}\sigma^{(M)}_{c\theta}\mathcal{R}_{c\alpha}(\nabla_{\theta}(\mathcal{R}^T)_{\alpha a})\notag\\
=&-r^{-1}\sigma^{(M)}_{c\theta}
\begin{pmatrix}
\cos\theta & \sin\theta \\
-\sin\theta & \cos\theta \\
\end{pmatrix}_{c\alpha}
\begin{pmatrix}
-\sin\theta & -\cos\theta \\
\cos\theta & -\sin\theta  \\
\end{pmatrix}_{\alpha a}\nonumber\\
=& r^{-1} \begin{pmatrix}
0 & -\sigma_{\theta\theta} \\
\sigma_{r\theta} & 0  \\
\end{pmatrix}.
\end{align}
Finally, we get
\begin{align}
\frac{\partial \sigma_{rr}}{\partial r}
+\frac{1}{r}(\sigma_{rr}-\sigma_{\theta\theta})
+f_{r}
=\rho\frac{\partial ^{2} u_{r}}{\partial t^{2}}\\
\frac{\partial \sigma_{r\theta}}{\partial r}
+2\frac{\sigma_{r\theta}}{r}
+f_{\theta}
=\rho \frac{\partial ^{2} u_{\theta}}{\partial t^{2}}\label{eq:sigma}.
\end{align}
By substituting Eq. (\ref{eq:s10}) into Eq. (\ref{eq:sigma}), we can get the equation of motion for $u_{\theta}$,
\begin{align}
\frac{\partial ^{2} u_{\theta}}{\partial r^{2}}
+\frac{1}{r}\frac{\partial u_{\theta}}{\partial r}
-\frac{1}{r^2}u_{\theta}
=\frac{2(1+\nu)}{E}\left(
\rho\frac{\partial^{2} u_{\theta}}{\partial t^{2}}-f_{\theta}\right),
\end{align}
where $f_{\theta}=\frac{\hbar}{2}\frac{\partial \dot{S}^{\prime}_{z}}{\partial r}$ is determined by spin evolution.
\section{Energy loss from damping}\label{loss of angular momentum and energy}
The energy losses resulting from displacement damping and spin damping need to be compensated by calculation. In Section \ref{Damped and undamped cases}, the drag force induced by $\eta$ is $F_{\mathrm{d}}^{\theta}=-\eta ~ {\partial u_{\theta}}/{\partial t}$. By utilizing the principle that energy power is equal to the product of force and velocity, we can estimate the energy loss as follows,
\begin{align}
	{E}_{1}=-\eta\int dt\int d^3 r(\frac{\partial u_{\theta}}{\partial t}\times\frac{\partial u_{\theta}}{\partial t}).
\end{align}
The energy loss directly brought about by spin damping $\zeta$ is solved like the solution of $E_{d}$ in Section \ref{Transfer of energy}, 
\begin{equation}
E_{2}=-Ia^2\int dt ~~ [\frac{\zeta}{\hbar S}(\boldsymbol{S}\times\boldsymbol{\Gamma})\times\boldsymbol{S}]\cdot\nabla^2\boldsymbol{S}.
\end{equation}
Besides, as the driving force is determined by the spin evolution, the energy of the displacement field is affected by the spin damping. This part of energy loss can be estimated in a similar manner as $E_1$,
\begin{equation}
E_{3}=\int dt \int d^3 r\frac{\zeta}{2S^{\prime}}(\frac{\partial }{\partial r}[(\boldsymbol{S}\times\Gamma)\times\boldsymbol{S}]_{z})\frac{\partial u_{\theta}}{\partial t}.
\end{equation}
\end{document}